\documentclass[aip,jcp,preprint]{revtex4-1}

\usepackage{graphics}
\usepackage{graphicx}
\usepackage{amsfonts}
\usepackage{textcomp}
\usepackage{amssymb}
\usepackage{amsmath}
\usepackage{rotating}
\usepackage{multirow}
\usepackage{makecell}
\usepackage{hyperref}

\begin{document}

\title{A Gaussian model of fluctuating membrane and its scattering properties}

\author{Cedric J. Gommes}  \email{cedric.gommes@uliege.be}
\affiliation{Department of Chemical Engineering, University of Li\`ege B6A, All\'ee du Six Ao\^ut 3, 4000 Li\`ege, Belgium}
\author{Purushottam S. Dubey} 
\affiliation{Forschungszentrum J\"ulich GmbH, J\"ulich Center for Neutron Science at the Heinz Maier Leibnitz Zentrum, Lichtenbergstrasse 1, 85747 Garching, Germany}
\author{Andreas M. Stadler}
\affiliation{Forschungszentrum J\"ulich GmbH, J\"ulich Center for Neutron Science, 52425 J\"ulich, Germany}
\affiliation{Institute of Physical Chemistry, RWTH Aachen University, Landoltweg 2, 52056 Aachen, Germany}
\author{Baohu Wu}
\affiliation{Forschungszentrum J\"ulich GmbH, J\"ulich Center for Neutron Science at the Heinz Maier Leibnitz Zentrum, Lichtenbergstrasse 1, 85747 Garching, Germany}
\author{Orsolya Czakkel}
\affiliation{Institut Laue-Langevin, 71 avenue des Martyrs, 38042 Grenoble Cedex 9, France}
\author{Lionel Porcar}
\affiliation{Institut Laue-Langevin, 71 avenue des Martyrs, 38042 Grenoble Cedex 9, France}
\author{Sebastian Jaksch}
\affiliation{Forschungszentrum J\"ulich GmbH, J\"ulich Center for Neutron Science at the Heinz Maier Leibnitz Zentrum, Lichtenbergstrasse 1, 85747 Garching, Germany}
\affiliation{European Spallation Source (ESS) ERIC, Partikelgatan 2, 224 84 Lund, Sweden}
\author{Henrich Frielinghaus}
\affiliation{Forschungszentrum J\"ulich GmbH, J\"ulich Center for Neutron Science at the Heinz Maier Leibnitz Zentrum, Lichtenbergstrasse 1, 85747 Garching, Germany}
\author{Olaf Holderer} \email{o.holderer@fz-juelich.de}
\affiliation{Forschungszentrum J\"ulich GmbH, J\"ulich Center for Neutron Science at the Heinz Maier Leibnitz Zentrum, Lichtenbergstrasse 1, 85747 Garching, Germany}

\date{\today}

\begin{abstract}
A mathematical model is developed, to jointly analyze elastic and inelastic scattering data of fluctuating membranes within a single theoretical framework. The model builds on a non-homogeneously clipped time-dependent Gaussian random field. This specific approach provides one with general analytical expressions for the intermediate scattering function, for any number of sublayers in the membrane and arbitrary contrasts. The model is illustrated with the analysis of small-angle x-ray and neutron scattering as well as with neutron spin-echo data measured on unilamellar vesicles prepared from phospholipids extracted from porcine brain tissues. The parameters fitted on the entire dataset are the lengths of the chain and head of the molecules that make up the membrane, the amplitude and lateral sizes of the bending deformations, the thickness fluctuation, and a single parameter characterizing the dynamics. 
\end{abstract}


\maketitle

\section{Introduction}


Membranes are ubiquitous in living systems, where they play a central role in controlling the interactions and exchanges of cells and organelles with their environment.\cite{Watson:2015} They are also common in synthetic systems, such as vesicles, lipo- or polymer-somes used as drug carriers or reaction compartments.\cite{Discher:2002,Guinart:2023} Developing analytical tools to investigate the nanometer-scale structure of membranes and their dynamics, is key to understanding their formation mechanisms, their physicochemical properties, and in the case of biological or neuronal membranes how they fulfill their function.\cite{Krugmann:2020, Krugmann:2021} In that general context, scattering methods (of either x-rays or neutrons) play a unique role because they enable one to characterize structures {\it in-situ} in their natural environment with nanometer resolution.\cite{Buldt:1978,Pabst:2010}

Scattering methods provide invaluable yet indirect structural and dynamical information in the form of correlation functions.\cite{VanHove:1954,Sivia:2011,Squires:2012,Glatter:2018,Luo:2023} A central aspect of any scattering investigation is, therefore, the development of suitable data analysis methods, by which the reciprocal-space data is converted to structurally significant information in real space. In that process, models are almost unavoidable.\cite{Pedersen:1997,Svergun:1999,Gommes:2018}

The simplest mathematical models of membranes describe them as a succession of flat layers with specific scattering-length densities. In that spirit, the main scattering characteristics of, say, a phospholipid membrane can be captured assuming a central hydrophobic layer squeezed between two hydrophilic layers. The description can be refined to account for more complex scattering-length density profiles, in relation {\it e.g.} to uneven water dissolution in the various segments,\cite{Kiselev:2008} or to the curvature of the membrane.\cite{Kiselev:2002,Pencer:2006,Chappa:2021} These types of models, however, do not capture the deformation of the actual membrane shape away from the ideal flat-layer structure.\cite{Monzel:2016} Such deformations contribute to a background scattering,\cite{Hamley:2022} and capturing them in a model would provide physical insights into the propensity of the membrane to bending and/or compression. Both deformation modes are central for a variety of physiological processes, including endo/exocytosis \cite{McMahon:2005} and cellular adhesion.\cite{Zidovska:2006}

More realistic membrane models can be produced through molecular dynamics simu\-lations.\cite{Venable:1993,Sodt:2014,Chakraborty:2020} This offers the possibility of building scattering data analyses on real physical interactions but the focus of this type of simulations is generally on molecular-scale processes, while the deformations of membranes are collective modes that involve thousands of molecules. Because of computational constraints, simulations are often restricted to small volumes, which is expected to overestimate the stiffness - {\it i.e.} underestimate the amplitude - of large-scale deformation modes. The accessible timescales are also in the picosecond range, which makes these models suitable for analyzing the diffusion of individual molecules. However, molecular-dynamic simulations of nanometer-scale deformations of membranes occurring over nanoseconds remain a challenge.\cite{Hayward:2022}

Interestingly, the models classically used to analyze the dynamics of membrane deformation through inelastic neutron scattering are distinctly different from those used to analyze the structure of the membrane through elastic scattering. When analyzing the data of neutron spin-echo experiments,\cite{Luo:2023} a central role is played by the theoretical results of Zilman and Granek concerning the bending fluctuations of two-dimensional films,\cite{Zilman:1996,Zilman:2002} whereby the time-dependence of the intermediate scattering function is described as stretched exponentials. These developments have been since generalized and adapted to other contexts,\cite{Nagao:2023,Granek:2024} also to characterize thickness fluctuations.\cite{Nagao:2009,Woodka:2012} With such models, however, the inelastic and elastic scattering are analyzed independently of one another, and this prevents one from building on the known structure of the membrane to understand its dynamics.

Here, we have developed a mathematical model of a fluctuating membrane, to jointly analyze elastic and inelastic scattering data within a single theoretical framework, with arbitrary sets of contrasts between the various phases that make up the membrane. The fluctuating parts of the model are captured through the statistics of a time-dependent Gaussian random field. Models built on clipped Gaussian fields have been used for many years to analyze scattering data from emulsions,\cite{Berk:1987,Berk:1991,Teubner:1991,Chen:1996} gels\cite{Roberts:1997,Gommes:2008} and porous materials.\cite{Levitz:1998,Prehal:2017,Gommes:2018} The present membrane modelling builds on three generalizations of the classical modelling approach. First, it uses a non-homogeneous clipping procedure to generate a structure with a more controlled morphology.\cite{Gommes:2009,Gommes:2020} It also allows for the presence of multiple regions with different scattering length densities.\cite{Gommes:2013} Finally, it allows the Gaussian field to be time-dependent in order to make the model suitable for inelastic scattering data analysis.\cite{Gommes:2021,Gommes:2022}

The structure of the paper is the following. The first section presents the experimental scattering data that motivate the theoretical development of the paper. For the purpose of comparison, a state-of-the-art analysis of the elastic and inelastic scattering data is presented. The Gaussian membrane model is presented in the following section, and a general expression is derived for the intermediate scattering function. The most technical parts of the derivations are presented in the Supporting Information. Finally, the discussion section is focused on the application of the model to analyze the scattering data presented in the experimental section.

\section{Experimental Section}
\label{sec:data}

\begin{figure}
\begin{center}
\includegraphics[width=7cm]{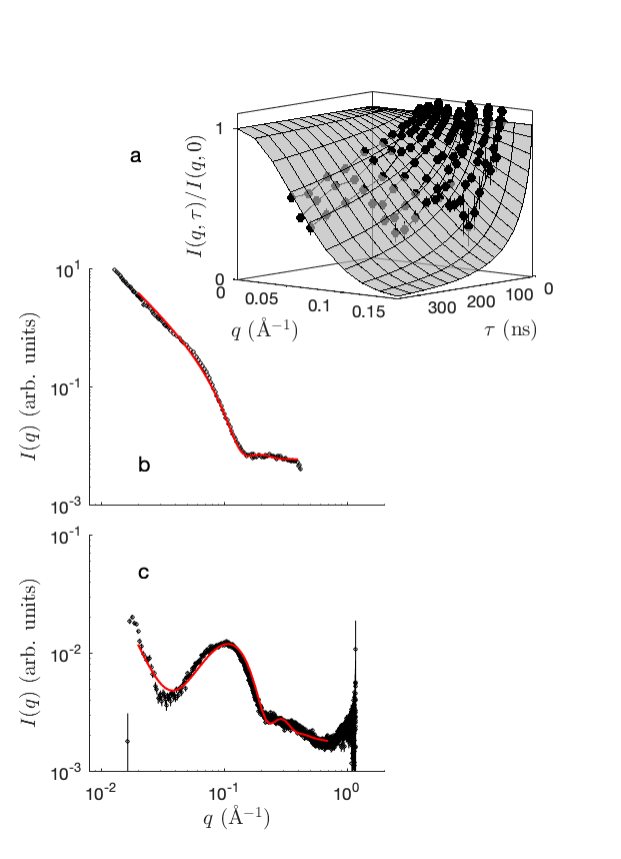}
\caption{\label{fig:data} Neutron Spin Echo (NSE) data of unilamellar vesicle (a), together with the corresponding Small-angle neutron (b) and X-ray (c) scattering patterns. The surface in (a) is a fit of the Zilman-Granek model, and the red lines in (b) and (c) result from the joint fitting of the SANS and SAXS data with a polydispersed three-layer slab model.}
\label{fig:data}
\end{center}
\end{figure}

The unilamellar vesicles (ULV) are prepared from phospholipids, extracted from porcine brain tissues as described by Folch et. al.\cite{Folch:1951} Figure \ref{fig:data} displays small-angle neutron (SANS) and X-ray(SAXS) scattering patterns measured on in-house SAXS instrument KWS-X (a Xeuss 3.0 XL from Xenocs), at the Forschungszentrum JÃŒlich, Garching and D22 SANS diffractometer at the Institut Laue-Langevin (ILL),\cite{illexperiment} respectively. The SANS measurements were performed at wavelength 11.5 \AA \ and 6 \AA \ and detector distances 17.6 m and 1.4 m, resulting in a $q$ range from 0.0026 to 0.64 \AA$^{-1}$. 

To provide a reference for the methodological developments of the paper, the SAXS and SANS data in Fig. \ref{fig:data} were first analyzed through a state-of-the-art method. A classical approach for analyzing small-angle scattering patterns of unilamellar vesicles, consists in assuming a spherical structure with radius much larger than the membrane thickness. In that case, the scattering can be approximated by the following separated form factor\cite{Kiselev:2002,Pencer:2006}
\begin{equation}
I (q) \simeq  \left|  4 \pi R^2 \frac{\sin(qR)}{qR} \int_{-\infty}^{+\infty} b(z) \cos(qz) \textrm{d}z \right|^2
\end{equation}
where $b(z)$ is the scattering-length density profile along a coordinate $z$ orthogonal to the membrane. This expression can be further simplified in the limit where $qR \gg 1$, {\it i.e.} when the $q$ range of interest concerns the inner structure of the membrane rather than the vesicle as a whole. Assuming any statistical distribution of the vesicle radius $R$, the scattering by the membrane becomes 
\begin{equation} \label{eq:I_profile}
I (q) \simeq     \frac{2 \pi A}{q^2}   \left| \int_{-\infty}^{+\infty} b(z) \cos(qz) \textrm{d}z \right|^2
\end{equation}
where $A = 4 \pi \langle R^2 \rangle$ is the average area of the vesicle.

The SANS and SAXS data in Fig \ref{fig:data} were fitted through Eq. (\ref{eq:I_profile}) to a three-layer scattering-length density profile, corresponding to two phospholipid molecules facing each other tail-to-tail. The neutron and x-ray scattering-length densities of the head and chain segments assumed for the fitting are reported in Tab. \ref{tab:slab_fit}. In practice, the logarithm of the SANS and SAXS intensities were fitted by least-square both independently and jointly, with the lengths of the chain and of the head $l_C$ and $l_H$ as fitting parameters. As shown in Tab. \ref{tab:slab_fit}, fitting the SANS and SAXS data independently from one another leads to incompatible values of the parameters, in particular for the head size $l_H$. Reasonable values of the parameters are obtained by minimizing jointly the error on both the SANS and SAXS. Polydispersity had to be assumed in the fit, to dampen the oscillations of the form factor that are not present in the data. This is quantified here as the standard deviation of the chain length relative to the mean; the value is $\sigma_C \simeq 15$ \%.
 
\begin{table}
\caption{Scattering-length densities relevant to neutron and x-ray scattering, and fitted parameters of the slab model in the case of SANS, SAXS and joint SANS/SAXS fitting. The errors are standard deviations of the fitted parameters observed through Monte Carlo simulations.}
\begin{center}
\begin{tabular}{lccccccc}
  &  $b_H$ & $b_C$ & $b_W$ & $l_H$ & $l_C$ & $\sigma_C$   \cr
  &  ($10^{-6}$ \AA$^{-2}$) &  ($10^{-6}$ \AA$^{-2}$)  & ( $10^{-6}$ \AA$^{-2}$)  &  (\AA) & (\AA) & (\%)  \cr 
  \hline
SANS &  1.87   &  -0.07  &  6.37  &   0.1 $\pm$ 0.0 & 19.3 $\pm$ 0.1   & 11.1 $\pm$ 1.1   \cr  
SAXS &  14.2  &  8.30  & 9.37   &   5.2 $\pm$ 0.0 & 18.7 $\pm$ 0.0   & 14.7 $\pm$ 0.6   \cr  
SANS \& SAXS &  {\it id.}  & {\it id.}  & {\it id.}   & 4.5 $\pm$ 0.4 & 17.1 $\pm$ 0.1  & 15.3 $\pm$ 0.3   \cr  
\end{tabular}
\end{center}
$b_{H/C/W}$: scattering length densities of the head and chain segments of the phospholipid, and of the outer water medium; $l_{H/C}$: length of the head and chain segments; $\sigma_C$: standard deviation of the chain length, relative to the average.
\label{tab:slab_fit}
\end{table}

Neutron spin-echo (NSE) spectroscopy is well suited to measure slow fluctuations of biological membranes such as phospholipid double layers.\cite{Nagao:2017, Nagao:2023} The intermediate scattering function $I(q,\tau)$, {\it i.e.} the Fourier transform of the real-space Van-Hove correlation function,\cite{Squires:2012} is measured on time scales up to several 100 ns on mesoscopic length scales (about 10-500 \AA). Neutron spin-echo measurements were carried out at the IN15 spectrometer at ILL,\cite{illexperiment} with three different wavelengths 12, 10 and 8\ \AA, with probing time scale up to 335 ns and $q$ range from 0.046 to 0.122 \AA$^{-1}$. 

The NSE data are shown in Fig. \ref{fig:data}a, under the form of the intermediate scattering function $I(q,\tau)$, normalized by the SANS intensity expressed in this context as  $I(q,0)$. To compare the developments of the paper with a classical approach, the data were fitted with the Zilman-Granek function\cite{Zilman:1996,Nagao:2023}, namely
\begin{equation} \label{eq:ZG}
\frac{I(q,\tau)}{I(q,0)} = \exp\left[ -B q^2 \tau^{2/3} \right]
\end{equation}
with parameter $B$ as the only fitting parameter for the entire dataset. The latter parameter can be given a physical interpretation in terms of the bending rigidity of the membrane but these values are notoriously dependent on the assumed viscosity of the medium.\cite{Nagao:2023} Assuming the viscosity of water ($\eta \simeq$ 10$^{-3}$ Pa.s), the value of $B$ extracted from the NSE data in Fig. \ref{fig:data}a converts to an effective bending modulus of 175 k$_B$T. 

\section{The Gaussian membrane model}

\subsection{Definition of the model}
\label{sec:model}

A classical approach to model disordered multiphasic structures consists in defining the various regions that make up a material based on whether the values taken locally by a given Gaussian random field is larger or smaller than a predefined threshold. \cite{Quiblier:1984,Berk:1987,Berk:1991,Teubner:1991,Lantuejoul:2002} In the context of scattering studies, this approach has notably been used to analyze elastic scattering from porous materials,\cite{Levitz:1998} polymer blends,\cite{Chen:1996} gels,\cite{Roberts:1997,Gommes:2008} confined liquids,\cite{Gommes:2013,Gommes:2018B} etc. Recently, this approach has been generalized to analyze inelastic scattering data as well through the use of time-dependent Gaussian fields.\cite{Gommes:2021}

\begin{figure}
\begin{center}
\includegraphics[width=12cm]{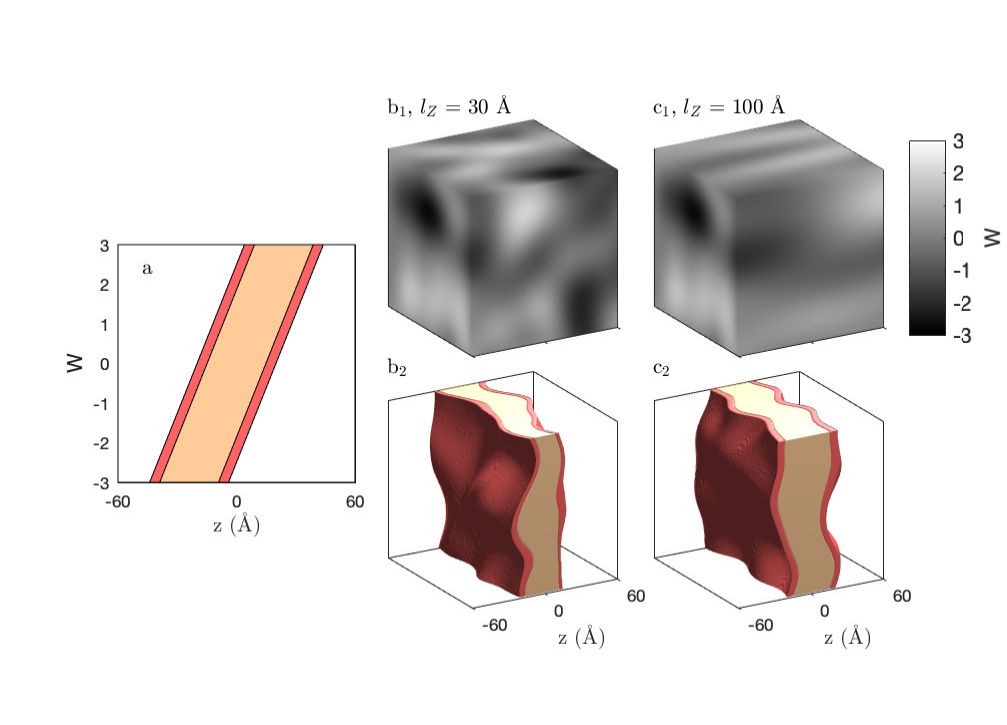}
\caption{\label{fig:sketch} Sketch of the membrane model, with (a) the model's flag (heads in red, chains in orange, and outer solution in white) and two Gaussian random fields having different correlation lengths $l_Z$ in direction $z$ (b$_1$ and c$_1$). In both cases, the correlation length in the orthogonal direction is $l_{XY} = 30$ \AA . The membrane structures resulting from these specific fields and flag are shown in b$_2$ and c$_2$.}
\end{center}
\end{figure}

The approach we propose here to model the structure and dynamics of fluctuating membranes is based on non-homogeneously clipped Gaussian fields,\cite{Gommes:2009,Gommes:2020} as sketched in Fig. \ref{fig:sketch}. In that spirit, the space- and time-dependent structure of the various regions that make up the membrane (head, chain, as well as outer solution) is modelled based on the combination of a stochastic Gaussian field $W(\mathbf{x},t)$ and of a set of deterministic rules used to convert the continuous values of the field into geometrical sets that describe the regions. In Fig. \ref{fig:sketch}, two Gaussian fields are shown in grey. The conversion rules are sketched in the colored flag, which can be thought of as a lookup table: whether one point $\mathbf{x}$ is assigned to any particular region at time $t$ depends on the position (here the distance $z$ to a given plane which controls the average orientation of the membrane) and on the local value of the random field $W(\mathbf{x},t)$. 

A Gaussian field is comprehensively characterized by its space- and time-dependent correlation function $g_W(\mathbf{r},\tau)$, which makes it a particularly useful concept in the context of scattering studies. The field correlation function is defined as
\begin{equation} \label{eq:gW_def}
g_W(\mathbf{r},\tau) = \langle W(\mathbf{x},t) W(\mathbf{x}+ \mathbf{r},t+\tau) \rangle
\end{equation}
where the brackets $\langle \rangle$ refer throughout the text to ensemble averages. In other words, $g_W(r,\tau)$ describes the statistical correlations between the values of $W(\mathbf{x},t)$ at two points at distance $\mathbf{r}$ from each other, with a time lag $\tau$.

A variety of procedures to construct Gaussian fields with specific space- and time-correlation functions are discussed for example by Gommes et al.\cite{Gommes:2021} The methods developed in the present paper are quite general and they apply to any of them. For the purpose of illustration, the two fields shown in the top row of Fig. \ref{fig:sketch} correspond to a correlation function of the type
\begin{equation} \label{eq:gW}
g_W(\mathbf{r}, 0)=  \exp \left[ - \frac{r_x^2 + r_y^2}{l_{XY}^2}  - \frac{r_z^2}{l_{Z}^2} \right] 
\end{equation}
where $r_x$, $r_y$, $r_z$ are the components of $\mathbf{r}$ parallel or perpendicular to the membrane; $l_{XY}$ and $l_Z$ are correlation lengths in those specific directions, which are two parameters that comprehensively describe the Gaussian field. In the cases of Figs. \ref{fig:sketch}b$_1$ and \ref{fig:sketch}c$_1$ the same correlation length is used in the $XY$ plane ($l_{XY}=30$ \AA), and the out-of plane correlation lengths are $l_{Z}=30$ \AA \ and 100 \AA, respectively. The correlation function in Eq. (\ref{eq:gW}) is written for the value $\tau=0$, and it characterizes therefore the spatial properties of an instantaneous snapshot of a possibly time-dependent field. To keep the discussion as general as possible, we postpone the discussion of specific time dependence to Sec. \ref{sec:time}. 

The limits between the layer boundaries are straight lines in the model's flag (Fig. \ref{fig:sketch}a). To keep the discussion general, we allow an arbitrary number of layers $N$ in the model, and we introduce $N+1$ linear threshold functions 
\begin{equation} \label{eq:clipping}
\alpha_n(z) = (z -Z_n)  / l_\alpha 
\end{equation}
with $n=0, \ldots N$, with parameters $Z_0 < Z_1< \ldots < Z_N$ and $l_\alpha$. Any point of space $\mathbf{x}$ is assigned to the $n^{th}$ layer at time $t$ if the value of the Gaussian field satisfies
\begin{equation} \label{eq:clipping_lamellae}
\alpha_{n}(z) \leq W(\mathbf{x},t) < \alpha_{n-1}(z)
\end{equation}
In the case of Fig. \ref{fig:sketch}, with $N=3$, the values are $Z_0=-24.5$ \AA, $Z_1=-16.5$ \AA, $Z_2=+16.5$ \AA, $Z_3=+24.5$ \AA,  and $l_\alpha= 10$\AA. The values of $Z_n$ control the most probable positions of the interfaces and the parameter $l_\alpha$ controls the slope of the boundary limits in the flag space, which converts to the amplitude of the interface fluctuations in real space. Flat interfaces centred exactly at $z=Z_n$ are obtained in the limiting case $l_\alpha \to 0$. 

In addition to the clipping parameters $l_\alpha$ and $Z_n$'s, the structure of the layers is also controlled by the field correlation function. In the particular case of Fig. \ref{fig:sketch} and Eq. (\ref{eq:gW}) the correlation length $l_{XY}$ controls the size of the bending-like patterns in the direction parallel to the membrane. The correlation length in the orthogonal direction $l_Z$ controls how the surface fluctuations at the different interfaces are correlated with each other. In the limit where $l_Z \gg |Z_{i} - Z_{j}|$ the values of the Gaussian field at $z = Z_i$ and $z=Z_j$ are strongly correlated ($g_W \simeq 1$) so that the fluctuations of the interfaces $i$ and $j$ are almost identical (Fig. \ref{fig:sketch} right). By contrast, if $l_Z \ll |Z_{i} - Z_{j}|$ the fluctuations of interfaces $i$ and $j$ are statistically independent of one another (Fig. \ref{fig:sketch} left). The value of $l_Z$ therefore controls the membrane thickness fluctuations. 

\begin{figure}
\begin{center}
\includegraphics[width=12cm]{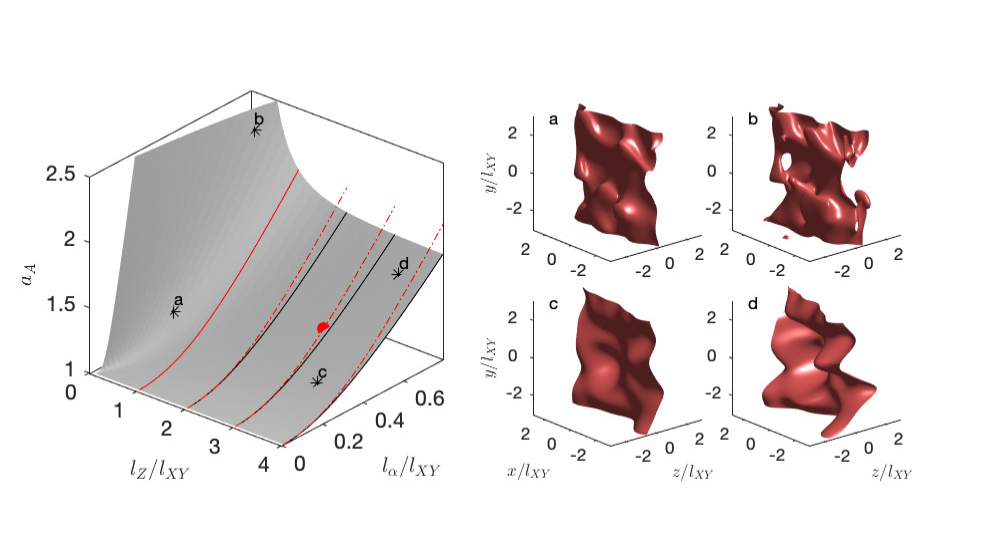}
\caption{\label{fig:area} Left panel: Dimensionless roughness factor $a_A$ of the interfaces in the membrane model, as a function of the field and clipping characteristic lengths. The solid red line at $l_Z/l_{XY}=1$ is the analytical result in Eq. (\ref{eq:roughness}), which is also plotted at larger values of $l_Z/l_{XY}$ to highlight the weak dependence on $l_Z$. Panels a to d point at specific realizations, also highlighted in the left panel. The red dot is the value inferred in Sec. \ref{sec:discussion} from the scattering data.}
\end{center}
\end{figure}

An important geometrical characteristic of the membrane is the specific surface area of the individual interfaces. This is conveniently expressed as a dimensionless roughness factor $a_A=a/A$, defined as the ratio of the area of the distorted interface $a$ to the area of the project flat surface $A$. A general expression for the roughness factor as a function of dimensionless numbers $l_Z/l_{XY}$ and $l_\alpha/l_{XY}$ is calculated in Sec. SI-I of the Supporting Information. The values are plotted in Fig. \ref{fig:area}, together with some realisations representative of various sets of parameters. In the particular case of an isotropic field, {\it i.e.} for $l_Z= l_{XY}$, the roughness can be calculated through a general expression derived in earlier work (see Eq. (6) of Ref. \onlinecite{Gommes:2009}) and expressed as follows
\begin{equation} \label{eq:roughness}
a_A = \frac{2}{\sqrt{\pi}} \frac{l_\alpha}{l_{XY}} \left\{ \exp\left[-\left(\frac{l_{XY}}{2 l_\alpha} \right)^2 \right] + \sqrt{\pi} \left(\frac{l_{XY}}{2 l_\alpha} + \frac{l_\alpha}{l_{XY}} \right) \textrm{erf}\left[ \frac{l_{XY}}{2 l_\alpha} \right] \right\}
\end{equation}
This expression is plotted in Fig. \ref{fig:area} as a solid red line. Globally, the dependence of $a_A$ on the correlation length $l_Z$ is strong only for $l_Z < l_{XY}$. This region corresponds to highly distorted and even disconnected interfaces ({\it e.g.} Fig. \ref{fig:area}b), which are unrealistic in the context of membranes. For parameters relevant to membranes, Eq. (\ref{eq:roughness}) provides a fair approximation of the roughness factor.

In addition to strictly geometrical aspects, the material characteristic relevant to scattering is the space- and time-dependent scattering-length density map $b(\mathbf{x},t)$, which quantifies the total scattering length in an infinitesimal volume centred on point $\mathbf{x}$ at time $t$. \cite{Sivia:2011,Squires:2012} If we assume that each layer $n$ in the membrane model has a specific composition-dependent scattering-length density $b_n$, the scattering length density map can be expressed as the following piecewise-constant function\cite{Gommes:2021}
\begin{equation} \label{eq:rho_piecewise}
b(\mathbf{x},t) = \sum_{n=0}^{N+1} b_n \mathcal{I}_n(\mathbf{x},t)
\end{equation}
where $\mathcal{I}_n(\mathbf{x},t)$ is the indicator function of layer $n$, which is equal to 1 if point $\mathbf{x}$ is in the $n^{th}$ layer at time $t$ and to 0 otherwise. Values of $b_n$ relevant to the head and chain segments of the phospholipid membrane in Fig. \ref{fig:data}, and the outer water solution are given in Tab. \ref{tab:slab_fit}. 

In the context of the non-homogeneously clipped Gaussian field of Eq. (\ref{eq:clipping_lamellae}), the indicator function of the $n^{th}$ layer is formally defined as
\begin{equation} \label{eq:I_n}
\mathcal{I}_n(\mathbf{x},t) = H\left[ W(\mathbf{x},t) - \alpha_n(z) \right] - H\left[ W(\mathbf{x},t) - \alpha_{n-1}(z) \right]
\end{equation}
where $H[]$ is Heaviside's step function (equal to 1 if the argument is positive and to 0 otherwise). The indicator functions of the outer regions ($n=0$ and $n=N+1$) are
\begin{equation} \label{eq:I_0}
\mathcal{I}_0(\mathbf{x},t) = H\left[ W(\mathbf{x},t) - \alpha_0(z) \right]
\end{equation}
and
\begin{equation} \label{eq:I_N+1}
\mathcal{I}_{N+1}(\mathbf{x},t) = 1 - H\left[ W(\mathbf{x},t) - \alpha_{N}(z) \right]
\end{equation}
corresponding to the water solution on both sides of the membrane. 

Note finally that although all illustrations ({\it e.g.} in Fig. \ref{fig:sketch}) assume a three-region model - with head, chain and water - of a single membrane, all the equations of the paper are general and the number of layers can be arbitrarily increased. This makes the present approach suitable also for refined models wherein the head and chain regions are decomposed into subregions with specific scattering-length densities reflecting possibly different water concentrations, \cite{Kiselev:2008,Lewis-Laurent:2021} or for multi-layer structures. \cite{Lemmich:1996,Krugmann:2020,Maiti:2021}

\subsection{Scattering properties}
\label{sec:scattering}

In the general context of stochastic models, the intermediate scattering function at wavevector $\mathbf{q}$ and time $\tau$ is calculated as the following Fourier transform\cite{Gommes:2021}
\begin{equation}
I(\mathbf{q},\tau) = \int dV_{1} \int dV_{2} \ e^{-i \mathbf{q} \cdot (\mathbf{x}_1-\mathbf{x}_2)}
 \langle b(\mathbf{x}_1,t)  b(\mathbf{x}_2,t+\tau)\rangle
\end{equation}
where the integrals are on $\mathbf{x}_1$ and $\mathbf{x}_2$, and the brackets $\langle \rangle$ stand for an ensemble average. In the particular case of a structure with a piecewise-constant scattering-length density like in Eq. (\ref{eq:rho_piecewise}), this can be written as
\begin{equation} \label{eq:I}
I(\mathbf{q},\tau) = \sum_{n=0}^{N+1} \sum_{m=0}^{N+1} b_n b_m P_{m,n}(\mathbf{q},\tau)
\end{equation}
with
\begin{equation} \label{eq:Pmn_3D}
P_{m,n}(\mathbf{q},\tau) = \int \textrm{d}V_1 \int \textrm{d}V_2 \quad e^{-i \mathbf{q} \cdot \left(\mathbf{x}_1 -  \mathbf{x}_2 \right) } S^{(2)}_{m,n}(\mathbf{x_1},\mathbf{x_2},\tau)
\end{equation}
In this expression $S^{(2)}_{m,n}(\mathbf{x_1},\mathbf{x_2},\tau)$ is a two-point correlation function, equal to the probability for point $\mathbf{x}_1$ to belong to layer $m$ at time $t$, and for $\mathbf{x}_2$ to belong to layer $n$ at later time $t + \tau$. In terms of the indicator functions, this is defined formally as
\begin{equation}
S^{(2)}_{m,n}(\mathbf{x_1},\mathbf{x_2}, \tau) = \langle \mathcal{I}_m (\mathbf{x}_1,t) \mathcal{I}_n (\mathbf{x}_2,t + \tau) \rangle
\end{equation}
As Gaussian random fields are second-order stationary,\cite{Lantuejoul:1991,Lantuejoul:2002} the time-dependence of the two-point functions in Eq. (\ref{eq:Pmn_3D}) is only through the difference $\tau=t_2-t_1$. Such simplification does not apply to the $\mathbf{x}_1$ and $\mathbf{x}_2$ dependence because the clipping procedure in Eq. (\ref{eq:I_n}) depends explicitly on the position $z$, so that the spatial stationarity of the Gaussian field is broken by the space-dependent clipping procedure\cite{Gommes:2009}.

In the limit where the time lag $\tau=t_2-t_1$ is much larger than any correlation time of the Gaussian field, the fluctuations at $(\mathbf{x}_1,t_1)$ and $(\mathbf{x}_2,t_2)$ are statistically independent. In that limit, the two-point correlation function simplifies to the following product
\begin{equation} \label{eq:S2_asymptotic}
\lim_{\tau \to \infty }S^{(2)}_{m,n}(\mathbf{x}_1,\mathbf{x}_2,\tau ) = S_m^{(1)}(\mathbf{x}_1) S_n^{(1)}(\mathbf{x}_2)
\end{equation}
where the one-point function
\begin{equation}
S_n^{(1)}(\mathbf{x}) = \langle \mathcal{I}^{(n)}(\mathbf{x},t) \rangle
\end{equation}
is the probability for point $\mathbf{x}$ to belong to layer $n$, at any given time. The limit in Eq. (\ref{eq:S2_asymptotic}) suggests decomposing the two-point correlation functions as follows
\begin{equation} \label{eq:S2_splitting}
S_{m,n}^{(2)}(\mathbf{x}_1, \mathbf{x}_1, \tau) = S^{(1)}_m(\mathbf{x}_1)S^{(1)}_m(\mathbf{x}_2) + \tilde S_{m,n}^{(2)}(\mathbf{x}_1, \mathbf{x}_1, \tau)  
\end{equation}
where the first term on the right-hand side is the time-independent contribution from the average structure, and $\tilde S_{m,n}^{(2)}$ is the contribution from the fluctuations.

The splitting of the two-point correlation function into average and fluctuating contributions, results in the following decomposition of the intermediate scattering function
\begin{equation} \label{eq:I_splitting}
I(\mathbf{q},\tau) = \bar I(\mathbf{q}) + \tilde I(\mathbf{q},\tau)
\end{equation}
where the two contributions $\bar I(\mathbf{q})$ and $\tilde I(\mathbf{q},\tau)$ are calculated from Eqs. (\ref{eq:I}) and (\ref{eq:Pmn_3D}) applied to the first and second terms on the right-hand side of Eq. (\ref{eq:S2_splitting}). We explicit these two contributions in the following two sections.

\subsubsection{Scattering contribution of the average structure}

The contribution of the average structure to the intermediate scattering function is calculated by applying Eqs. (\ref{eq:I}) and (\ref{eq:Pmn_3D}) to the first term in Eq. (\ref{eq:S2_splitting}). This results in
\begin{equation} \label{eq:bar_I}
\bar I(\mathbf{q}) = \left| \int e^{-i \mathbf{q} \cdot \mathbf{x}} \ \bar b(\mathbf{x}) \ \textrm{d}V_x \right|^2
\end{equation}
where 
\begin{equation} \label{eq:b_bar}
\bar b(\mathbf{x}) = \sum_{n=0}^{N+1} b_n S_n^{(1)}(\mathbf{x})
\end{equation}
is the average scattering-density map. Equation (\ref{eq:bar_I}) is classical, and it reduces to Eq. (\ref{eq:I_profile}) in the case of a large vesicle. It has to be stressed, however, that in the present context, $\bar b(z)$ is not the true instantaneous scattering-length density profile, but its average value smoothed over time.

\begin{figure}
\begin{center}
\includegraphics[width=5cm]{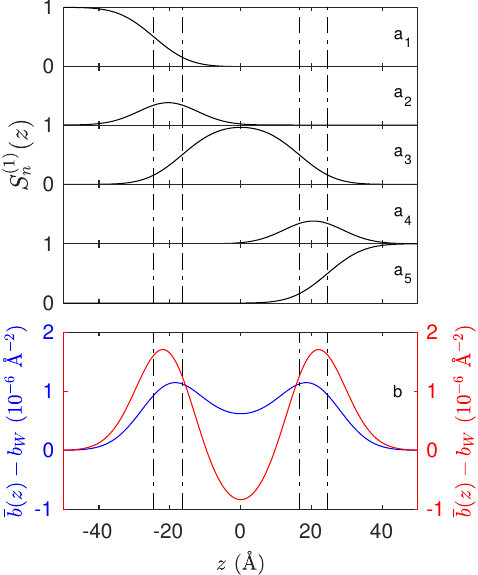}
\caption{\label{fig:S1} One-point probabilities $S_n^{(1)}(z)$ of the various layers of the membrane model in Fig. \ref{fig:sketch} (a$_1$ to a$_5$), and (b) corresponding average scattering-length density profile relevant to x-ray (red) and neutron (blue) scattering. The vertical dashed lines are the most probable positions of the water/head and head/chain interfaces.}
\end{center}
\end{figure}

In the case of the membrane model in Eq. (\ref{eq:I_n}), the one-point probability function of layer $n$ is calculated as
\begin{equation} \label{eq:S_n}
S^{(1)}_n(\mathbf{x}) = \Lambda_1[\alpha_{n}(z)] - \Lambda_1[\alpha_{n-1}(z)] 
\end{equation}
where $\Lambda_1[\alpha]$ is the probability for a centred Gaussian variable with unit variance to take values larger than $\alpha$. This can also be written as
\begin{equation} \label{eq:Lambda1}
\Lambda_1[\alpha] = \frac{1}{2} \left( 1 - \textrm{erf} \left[ \frac{\alpha}{\sqrt{2} } \right] \right)
\end{equation}
where erf[] is the error function. The one-point functions of the two outer regions are 
\begin{eqnarray} \label{eq:phi}
S^{(1)}_0(\mathbf{x},t) &=& \Lambda_1[\alpha_0(z)] \cr
S^{(1)}_{N+1}(\mathbf{x},t) &=& 1- \Lambda_1[\alpha_{N}(z)] 
\end{eqnarray}
The one-point probability functions corresponding to Fig. \ref{fig:sketch} are illustrated in Fig. \ref{fig:S1}, together with the average scattering-length density profiles relevant to x-ray and neutron scattering. 

The average scattering-length density profiles do not depend on any characteristic of the Gaussian field, so that the profiles are identical for the compressible and incompressible cases shown in Figs. \ref{fig:sketch}b$_2$ and  \ref{fig:sketch}c$_2$. The corresponding scattering contribution $\bar I(q)$ relevant to x-ray and neutron are plotted as black lines in Fig. \ref{fig:C_and_I}c and Fig. \ref{fig:C_and_I_neutron}c, respectively. Because $\bar I(q)$ is the Fourier transform of a smooth function, with no singularity in any of its derivatives, it decreases asymptotically with $q$ faster than any power law.\cite{Lighthill:1958} In particular, with the splitting of the scattering in Eq. (\ref{eq:I_splitting}) it is the fluctuation contribution that is responsible for Porod's $q^{-4}$ scattering.

\subsubsection{Scattering by the fluctuations}
\label{sec:fluctuations}

\begin{figure}
\centering
\includegraphics[width=5cm]{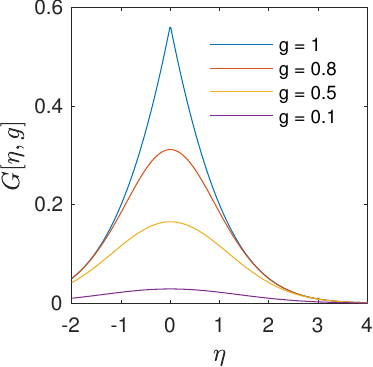}
\caption{Function $G[\eta,g]$ defined in Eq. (\ref{eq:G}) }
\label{fig:Gamma_L}
\end{figure}

The contribution of the fluctuations to the scattering $\tilde I(\mathbf{q},\tau)$ is specific to the present stochastic model, and it has no equivalent in the deterministic slab model of Eq. (\ref{eq:I_profile}). It accounts for the time-dependent bending-like deformation of the membrane as well as for its thickness fluctuations. Mathematically, it is calculated by applying Eqs. (\ref{eq:I}) and (\ref{eq:Pmn_3D}) to the second term in Eq. (\ref{eq:S2_splitting}). We show in Sec. SI-II of the Supporting Information that its value per unit area $A$ of the membrane is calculated as the Fourier transform of the following correlation function
\begin{eqnarray} \label{eq:C_tilde}
\tilde C_{b}(\mathbf{r},\tau) &=& \left[ \sum_{n=0}^N (b_n - b_{n+1})^2 \right]  \Gamma_{0}(\mathbf{r},\tau) \cr
&+&  \sum_{m > n} (b_n - b_{n+1})(b_m - b_{m+1}) \left[  \Gamma_{Z_m-Z_n}(\mathbf{r},\tau) + \Gamma_{Z_n-Z_m}(\mathbf{r},\tau) \right]
\end{eqnarray}
where the function $ \Gamma_{L}(\mathbf{r},\tau)$ is given by 
\begin{equation} \label{eq:Gamma_mn}
\Gamma_{L} (\mathbf{r},\tau) = l_\alpha G\left[ \frac{|r_z-L| }{l_\alpha}, g_W\left( \mathbf{r}, \tau \right) \right]
\end{equation}
with
\begin{eqnarray} \label{eq:G}
G[\eta,g] &=& \frac{1}{\sqrt{\pi}} \left(\exp\left[-\frac{\eta^2}{4} \right] - \exp\left[-\frac{\eta^2}{4(1-g)} \right] \sqrt{1-g}  \right) \cr
&+& \frac{\eta}{2} \left( \textrm{erf}\left[ \frac{\eta}{2} \right] -   \textrm{erf}\left[ \frac{\eta}{2 \sqrt{1-g}} \right] \right)
\end{eqnarray}
The function $G[\eta,g]$ is plotted in Fig. \ref{fig:Gamma_L} against its two arguments $\eta$ and $g$. Note that in the particular case where $l_\alpha$ becomes vanishingly small, the stochastic model reduces to a deterministic slab model. In that case, all $\Gamma$'s in Eq. (\ref{eq:C_tilde}) converge to zero, and the only contribution left to the scattering is from the average structure $\bar I(q)$, as it should.

\begin{figure}
\centering
\includegraphics[width=7cm]{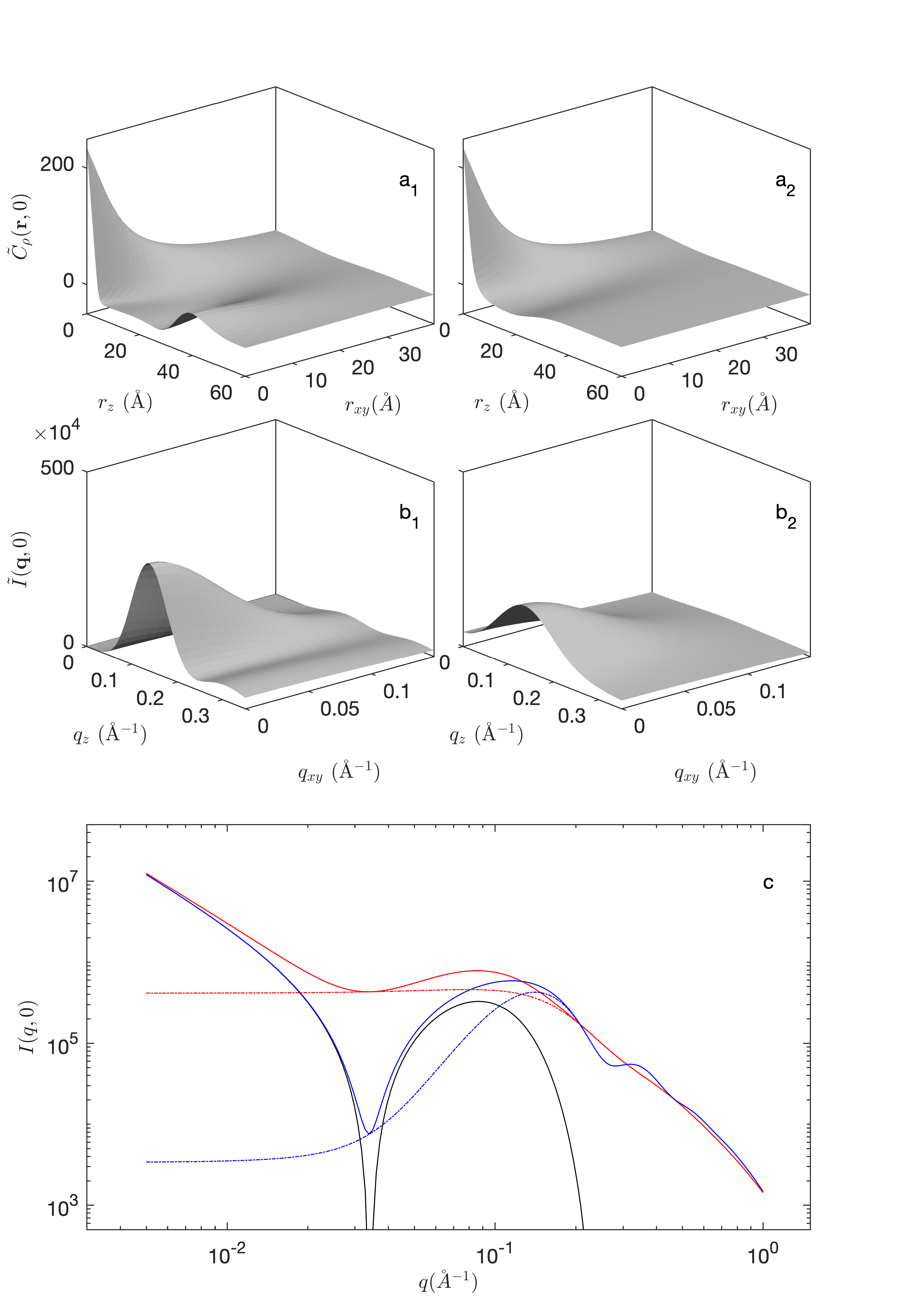}
\caption{\label{fig:C_and_I} Fluctuation contribution to the correlation function $\tilde C_b(\mathbf{r},0)$ (top) and scattered intensity $\tilde I(\mathbf{q},0)$ (middle), together with the rotationally-averaged scattered intensity (bottom), in the case of x-ray contrast. The parameters are those relevant to the membrane models in Fig. \ref{fig:sketch}, with $l_{XY} = 30$ \AA \ and $l_Z = 30$ \AA \ (a$_1$, b$_1$) or $l_Z = 100$ \AA \ (a$_2$, b$_2$). In c, the black line is the scattering from the average structure. The colored dashed and solid lines are the fluctuation contributions and the total scattering, for $l_Z = 30$ \AA \ (red) and $l_Z = 100$ \AA \ (blue).}
\end{figure}

Figures \ref{fig:C_and_I}a$_1$ and  \ref{fig:C_and_I}a$_2$ display the correlation functions $\tilde C_b(\mathbf{r},0)$ relevant to the two membrane models in Fig. \ref{fig:sketch}, as a function of the in-plane and out-of-plane components of $\mathbf{r}$, namely $r_{xy}$ and $r_z$. The x-ray contrast is assumed for the figure. The $r_{xy}$ dependence of the correlation function, corresponding to in-plane vectors $\mathbf{r}$, are very similar for the two models. It characterises the wavelike bending of the membrane with a typical size $l_{XY} = 30$ \AA \ in the case of the figure, independently of the out-of-plane correlation length $l_Z$. The $r_z$ dependences, however, are distinctly different in Figs. \ref{fig:C_and_I}a$_1$ and \ref{fig:C_and_I}a$_2$. Strong oscillations appear in $\tilde C(\mathbf{r},0)$ for $l_Z = 100$ \AA, because in that case the membrane bends while keeping a constant thickness (Fig. \ref{fig:sketch} right). By contrast, the membrane thickness fluctuates in the case with $l_Z = 30$ \AA \ (Fig. \ref{fig:sketch} left) which leads to blunter features in the correlation function. 

The 3D Fourier transforms $\tilde I(\mathbf{q},0)$ exhibit the same differences in reciprocal space. They are shown in Figs. \ref{fig:C_and_I}b$_1$ and  \ref{fig:C_and_I}b$_2$ against in-plane and out-of-plane components of the scattering vector $q_{xy}$ and $q_z$. Because of the cylindrical symmetry of the system, they were calculated as
\begin{equation}
\tilde I(q_{xy},q_z,\tau) =  4\pi \int_0^\infty \textrm{d}r_z \int_0^\infty r_{xy} \textrm{d}r_{xy} \  J_0(q_{xy} r_{xy}) \cos(q_z r_z) \tilde C_b(r_{xy},r_z,\tau)
\end{equation}
where $J_0()$ is the Bessel function of order 0. The rotationally-averaged intensity, relevant to SANS and SAXS analysis, is then obtained as
\begin{equation} \label{eq:I_q_tau}
\tilde I (q,\tau) = \int_0^{\pi/2} \sin(\theta) \textrm{d}\theta \ \tilde I(q \sin(\theta),q \cos(\theta),\tau)
\end{equation}
The so-obtained value of $\tilde I(q,0)$ are shown in Fig. \ref{fig:C_and_I}c. 

The membrane thickness fluctuation is found to have two distinctly different effects on the scattering patterns. First and foremost, it enhances the scattering at low $q$. This is visible in the different values of the low-$q$ plateaus in the fluctuation contributions $\tilde I(q,0)$ in Fig. \ref{fig:C_and_I}, which is about 100 times smaller for the larger value of  $l_Z$. This effect is physically identical to the well-known enhancement of forward scattering by compressibility\cite{Glatter:2018}. As a second, weaker, effect: thickness fluctuations add polydispersity to the distance between the hydrophilic heads on both sides of the membrane, and this dampens the oscillations at intermediate $q$. By contrast, the high-$q$ scattering is almost blind to thickness fluctuations, because it is dominated by surface-dependent $q^{-4}$ scattering and the areas are only weakly dependent on $l_Z$ (see Fig. \ref{fig:area}). 

\begin{figure}
\centering
\includegraphics[width=7cm]{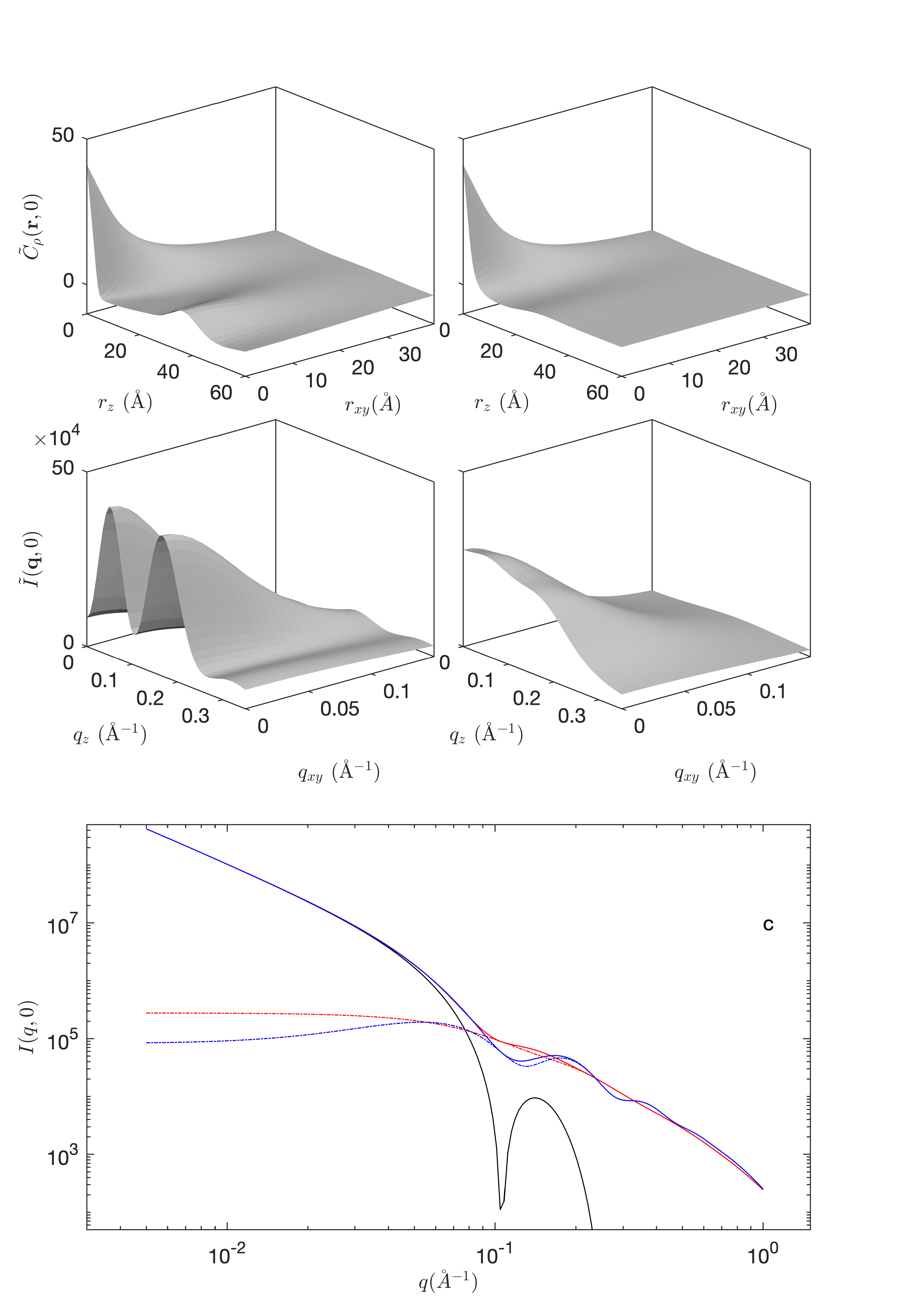}
\caption{\label{fig:C_and_I_neutron} Fluctuation contribution to the correlation function $\tilde C_b(\mathbf{r},0)$ (top) and scattered intensity $\tilde l(\mathbf{q},0)$ (middle), together with the rotationally-averaged scattered intensity (bottom), in the case of neutron contrast. The parameters are those relevant to the membrane models in Fig. \ref{fig:sketch}, with $l_{XY} = 30$ \AA \ and $l_Z = 30$ \AA \ (a$_1$, b$_1$) or $l_Z = 100$ \AA \ (a$_2$, b$_2$). In c, the black line is the scattering from the average structure. The colored dashed and solid lines are the fluctuation contributions and the total scattering, for $l_Z = 30$ \AA \ (red) and $l_Z = 100$ \AA \ (blue).}
\end{figure}

Despite the strong effect of thickness fluctuations on forward scattering, one has to consider also that the total scattering at low $q$ is dominated by the average-structure contribution. This is manifest through Eq. (\ref{eq:I_profile}), which reduces to
\begin{equation}
\bar I(q) \simeq \frac{2 \pi A}{q^2} \times \left| \int_{-\infty}^{+\infty} \bar b(z) dz \right|^2
\end{equation}
at low $q$, where the second factor is the squared total excess scattering length of the membrane. The latter $q^{-2}$ scattering hides the compressibility-related forward scattering, which becomes visible only at intermediate $q$. This effect is much stronger for neutron-like chain contrast (Fig. \ref{fig:C_and_I_neutron}) than for x-ray like head contrast (Fig. \ref{fig:C_and_I}).

\subsubsection{Inelastic scattering by time-dependent fluctuations}
\label{sec:time}

In all mathematical expressions discussed in Sec. \ref{sec:fluctuations}, the dependence of the scattering intensity on the Gaussian field is through the field correlation function $g_W$. This is notably the case for the fluctuation contribution to the scattering in Eqs. (\ref{eq:C_tilde}) and (\ref{eq:Gamma_mn}), which are valid whether the Gaussian field is static or time-dependent. In the latter case, the field correlation function depends on the time lag $\tau$ in line with its definition in Eq. (\ref{eq:gW_def}), and so does the intermediate scattering function $I(q,\tau)$. We consider now four different time-dependent Gaussian random fields, all of which correspond to the same spatial statistics as the correlation function $g_W(\mathbf{r},0)$ in Eq. (\ref{eq:gW}).

When it comes to adding time dependence to a static Gaussian random field, it is convenient to think of it as a superposition of a large number of elementary wave packets $w(\mathbf{x})$ localized in space, namely
\begin{equation} \label{eq:W_sum}
W(\mathbf{x}) =\sum_s A_s w(\mathbf{x} -  \mathbf{x}_s)
\end{equation}
where the points $\mathbf{x}_s$ are uniformly distributed in space, according to a Poisson point process with density $\theta$, and $A_s$ are random and statistically independent amplitudes. The so-constructed field has correlation function \cite{Lantuejoul:2002,Gommes:2021}
\begin{equation} \label{eq:gW_sum}
g_W(\mathbf{r},0) = \theta \langle A^2 \rangle K(\mathbf{r})
\end{equation}
where
\begin{equation}
K(r) = \int \textrm{d}V_x \ w(\mathbf{x}) w(\mathbf{x} - \mathbf{r})
\end{equation}
is the self-convolution of the elementary wave packet.

In the limit of infinitely large density of waves $\theta$, the value of $W(\mathbf{x})$ is the sum of a large number of independent contributions. It then results from the central-limit theorem that the local values of the field are Gaussian distributed. Moreover, as a consequence of Eq. (\ref{eq:gW_sum}), one has to impose $\theta \langle A^2 \rangle K(0) = 1$ to ensure that the field has variance equal to one. The specific wave corresponding to the field correlation function in Eq. (\ref{eq:gW}) is
\begin{equation}
w(\mathbf{x}) = \exp\left[ -2 \left( \frac{x^2+y^2}{l_{XY}^2} + \frac{z^2}{l_Z^2} \right)\right]
\end{equation}
with self-convolution
\begin{equation} \label{eq:K_wave}
K(\mathbf{r}) = \frac{\pi^{3/2} l_{XY}^2 l_Z}{8}  \exp\left[ - \left( \frac{r_x^2+r_y^2}{l_{XY}^2} + \frac{r_z^2}{l_Z^2} \right)\right]
\end{equation}
A variety of other wave shapes and their corresponding field correlation functions are given in Tab. 1 of Ref. \onlinecite{Gommes:2021}.
 
In the first type of time-dependent models we consider, the amplitudes $A_s$ of the waves are let to oscillate with frequency $\omega_s$ and phase $\varphi_s$, namely
\begin{equation}
W(\mathbf{x},t) =\sqrt{2} \sum_s A_s w(\mathbf{x}-\mathbf{x}_s) \cos(\omega_s t - \varphi_s)
\end{equation}
where the factor $\sqrt{2}$ keeps the variance of the field equal to one. The phases are chosen uniformly in $[0, 2\pi)$, and the frequencies are distributed according to any user-specified spectral density $f(\omega) \textrm{d}\omega$.  With such a construction, the space- and time-dependent correlation function takes the following separable form \cite{Gelfand:2010,Gommes:2021}
\begin{equation} \label{eq:g_separable}
g_W(\mathbf{r},\tau) = g_W(\mathbf{r},0) g'(\tau)
\end{equation}
where the time-dependent factor depends on the spectral density as follows
\begin{equation}
g'(\tau) =\int_0^\infty \cos[\omega \tau] f'(\omega) \textrm{d}\omega
\end{equation}
This cosine transform can in principle be inverted to calculate the spectral density corresponding to any desired correlation function. Not all correlation functions, however, are realizable as some may lead to non-positive spectral densities.

\begin{figure}
\centering
\includegraphics[width=8cm]{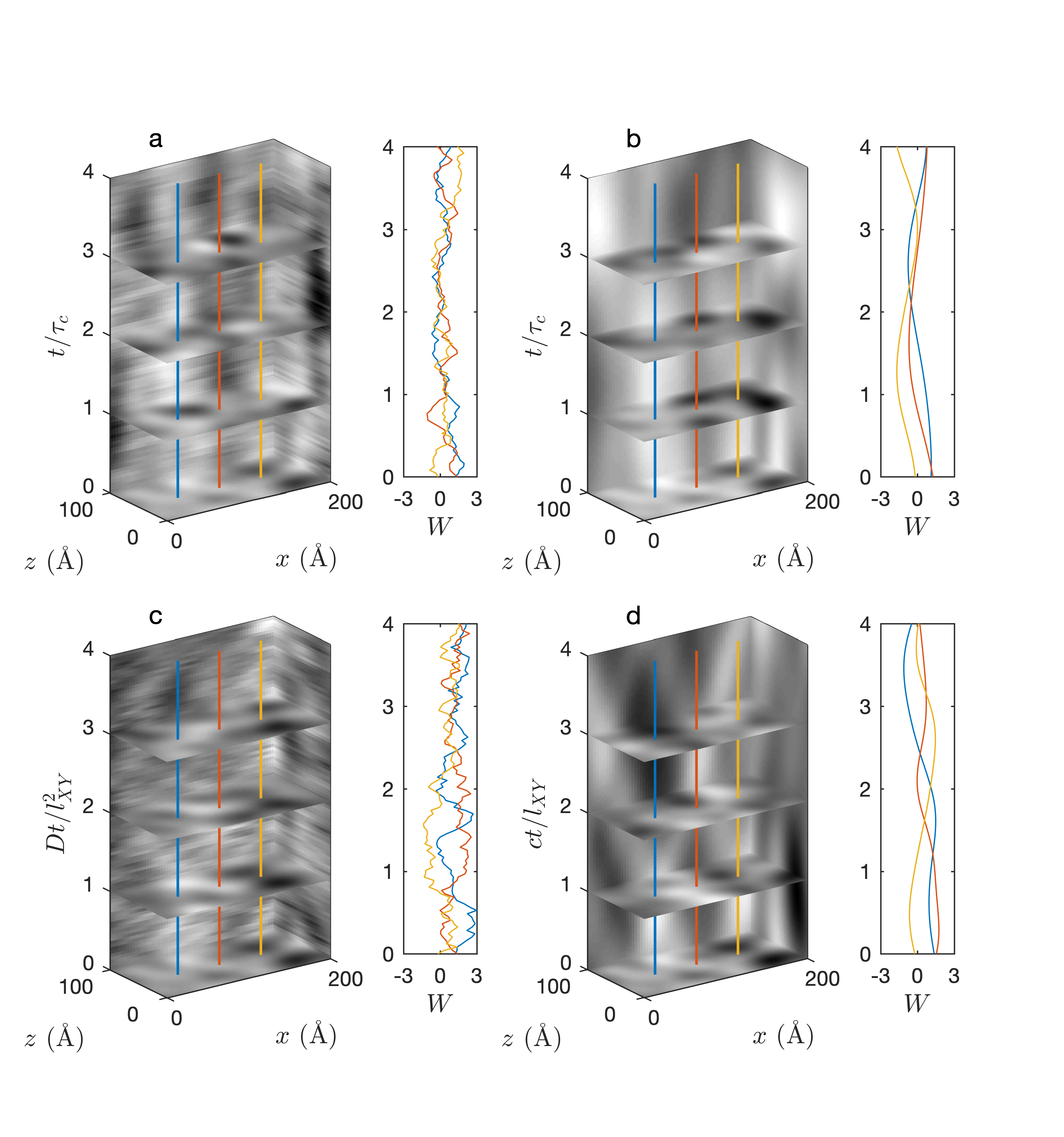}
\caption{Realizations of time-dependent Gaussian fields with $l_{XY}=l_{Z}=30$ \AA, and various time dependences: (a and b) exponential and hyperbolic secant correlation with correlation time $\tau_c$; (c) diffusive wave motion with diffusion coefficient $D$; (d) ballistic wave motion with velocity $c$. In each case, the values of the field $W$ sampled at three points are plotted against time. Note how the two models on the right are differentiable with respect to time, and the models on the left are not.}
\label{fig:GRF_time}
\end{figure}

For modelling purposes, a natural choice is an exponential correlation $g'(\tau) = \exp(-\tau/\tau_c)$ with correlation time $\tau_c$ as a parameter. This corresponds to the following spectral density
\begin{equation} \label{eq:f_exp}
f'(\omega) = \frac{1}{\pi} \frac{2 \tau_c}{1+(\omega \tau_c)^2}
\end{equation}
which is indeed positive and realizable. A realization of this field is shown in Fig. \ref{fig:GRF_time}a, together with the time dependent values sampled at three different points. This specific field is not differentiable with respect to time, which is a direct consequence of the linearity of $g'(\tau)$ for small $\tau$.\cite{Gommes:2021} A smoother dynamics - with a differentiable field - can be obtained by choosing a time correlation of the type $g'(\tau) = 1/\cosh[\tau/\tau_c]$. This function is identical to an exponential for large times, but it is quadratic for small values of $\tau$. It corresponds to the spectral density
\begin{equation}
f'(\omega) =\frac{2 \tau_c \cosh[\pi \omega \tau_c/2]}{1+\cosh[ \pi \omega \tau_c]}
\end{equation}
A realization of this field is shown in Fig. \ref{fig:GRF_time}b. Its differentiability is apparent when looking at the values sampled at specific points in space. 

An alternative approach for creating time-dependent Gaussian fields starting from the wave superposition in Eq. (\ref{eq:W_sum}), consists in keeping the amplitudes of the waves $A_s$ constant but letting the waves move in space with specific probabilistic laws, namely
\begin{equation}
W(\mathbf{x},t) = \sum_s A_s w(\mathbf{x}-\mathbf{x}_s - \mathbf{j}_s(t))
\end{equation}
where $\mathbf{j}_s(t)$ is the displacement jump of wave $s$ at time $t$, away from its initial position $\mathbf{x}_s$. The field correlation function corresponding to this construction is \cite{Gommes:2021}
\begin{equation} \label{eq:gW_jump}
g_W(\mathbf{r},\tau) = \theta \langle A^2 \rangle \int \textrm{d}V_j \ K(\mathbf{r} - \mathbf{j}) f_\tau(\mathbf{j})
\end{equation}
where $f_\tau(\mathbf{j}) \textrm{d}V_j$ is the jump distribution probability over a duration $\tau$. 

In the present context of membrane modelling with the clipping procedure in Fig. \ref{fig:sketch}, it is only the waves close to $z=0$ that contribute to the structure fluctuations. Accordingly, we focus on two-dimensional displacements of elementary waves within the $xy$ plane. We consider two distinctly different in-plane displacement laws. The first is diffusive, and it corresponds to a jump probability \cite{Berg:1993}
\begin{equation}
f_\tau(\mathbf{j}) = \left( 4 \pi D \tau \right)^{-1} \exp \left[ - \frac{j_x^2+j_y^2}{4 D \tau} \right] \delta(j_z)
\end{equation}
where $D$ is a diffusion coefficient, which parameterizes the dynamics. In this equation, the exponential corresponds to a two-dimensional random walk, the Dirac function $\delta(j_z)$ restricts the motion to being parallel to $xy$, and the first factor normalizes the probabilities to one. According to Eq. (\ref{eq:gW_jump}), this specific displacement law leads to field correlation function
\begin{equation} \label{eq:g_diffusive}
g_W(\mathbf{r},\tau) = \left(1 + \frac{4 D \tau}{l_{XY}^2} \right)^{-1} \exp\left[ -\frac{r_{xy}^2}{4 D \tau + l_{XY}^2} \right] \exp\left[ -\frac{r_{z}^2}{l_{Z}^2}  \right]
\end{equation}
where we have used the values of $K(\mathbf{r})$ from Eq. (\ref{eq:K_wave}). Note that this time-dependent correlation function reduces to Eq. (\ref{eq:gW}) for $\tau =0$, as it should. A particular realization of this Gaussian field is shown in Fig. \ref{fig:GRF_time}c. This field is also non-differentiable with respect to time because its correlation function $g_W(0,\tau)$ in Eq. (\ref{eq:g_diffusive}) is linear in $\tau$ at the origin.

A smooth time-differentiable field can be generated with moving waves assuming their ballistic motion with velocity $c$ and random directions parallel to the $xy$ plane. This corresponds to the following jump probability distribution
\begin{equation}
f_\tau(\mathbf{j}) = \frac{\delta(j_{xy}- c \tau)}{2 \pi j_{xy}} \delta(j_z)
\end{equation}
where $j_{xy}=\sqrt{j_x^2 + j_y^2}$ and $j_z$ are the in-plane and out of plane displacements. Here too the delta function restricts the motion to being parallel to $xy$, and the denominator $2 \pi j_{xy}$ normalizes the probabilities to one. The field correlation function calculated from Eq.(\ref{eq:gW_jump}) is 
\begin{equation}
g_W(\mathbf{r},\tau) =  \exp\left[ -\frac{r_{z}^2}{l_{Z}^2}  \right] \exp\left[ -\frac{(r_{xy}-c \tau)^2}{l_{XY}^2} \right]  \exp\left[ - \frac{2 r_{xy} c \tau }{l_{XY}^2} \right]
I_0\left[ \frac{2 r_{xy} c \tau }{l_{XY}^2} \right]
\end{equation}
where $I_0()$ is the modified Bessel function of the first kind. This expression of $g_W(\mathbf{r},\tau)$ also reduces to Eq. (\ref{eq:gW}) for $\tau =0$. A realization of the ballistic field is shown in Fig. \ref{fig:GRF_time}d.

\begin{figure}
\centering
\includegraphics[width=8cm]{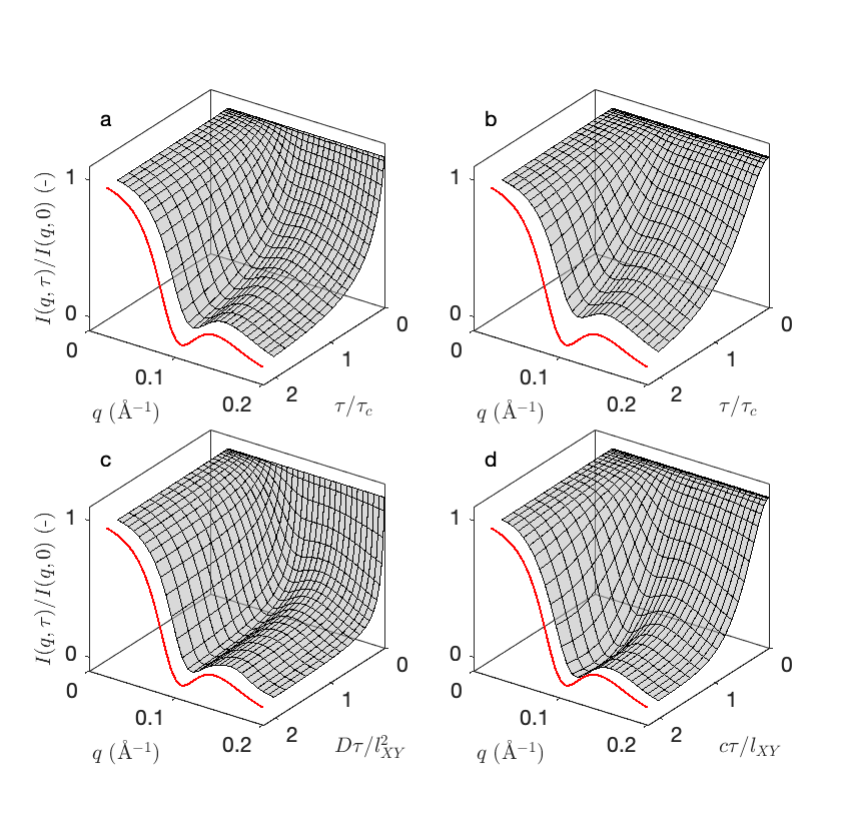}
\caption{Intermediate scattering functions calculated from the parameters of Fig. \ref{fig:sketch} ($l_{Z}=30$ \AA) and various time dependences: (a and b) exponential and hyperbolic secant correlation with correlation time $\tau_c$; (c) diffusive wave motion with diffusion coefficient $D$; (d) ballistic wave motion with velocity $c$. In all cases, the solid red line is the scattering from the average structure. Note the qualitatively different time dependence of $I(q,\tau)$ for large $q$ and small $\tau$ in the left and right graphs.}
\label{fig:Intermediate}
\end{figure}

The intermediate scattering functions calculated from Eq. (\ref{eq:I_q_tau}) with the same parameters as in Fig. \ref{fig:sketch} are plotted in Fig. \ref{fig:Intermediate} for the four specific types of dynamics considered in Fig. \ref{fig:GRF_time}. In all cases, the intermediate scattering functions - expressed as $I(q,\tau)/I(q,0)$ - decrease towards $\bar I(q)/I(q,0)$ for large values of $\tau$, where $\bar I(q)$ is the scattering by the average structure, calculated through Eq. (\ref{eq:bar_I}). Mathematically, this results from the fact that the fluctuation contribution to scattering $\tilde I(q,\tau)$ vanishes when the field correlation $g_W$ is equal to zero, as is asymptotically the case for $\tau \to \infty$.  The two types of non-differentiable or smooth time-dependences identified in Fig. \ref{fig:GRF_time}, lead to qualitatively different shapes of the intermediate scattering function at high $q$ and low $\tau$. Smooth dynamics lead to quadratic $\tau$ dependence that does not seem to be observed experimentally.\cite{Gommes:2021} Other differences between the various dynamics are discussed later, when using the models to fit the NSE data from Fig. \ref{fig:data}.

\section{Discussion}
\label{sec:discussion}

\begin{figure}
\centering
\includegraphics[width=8cm]{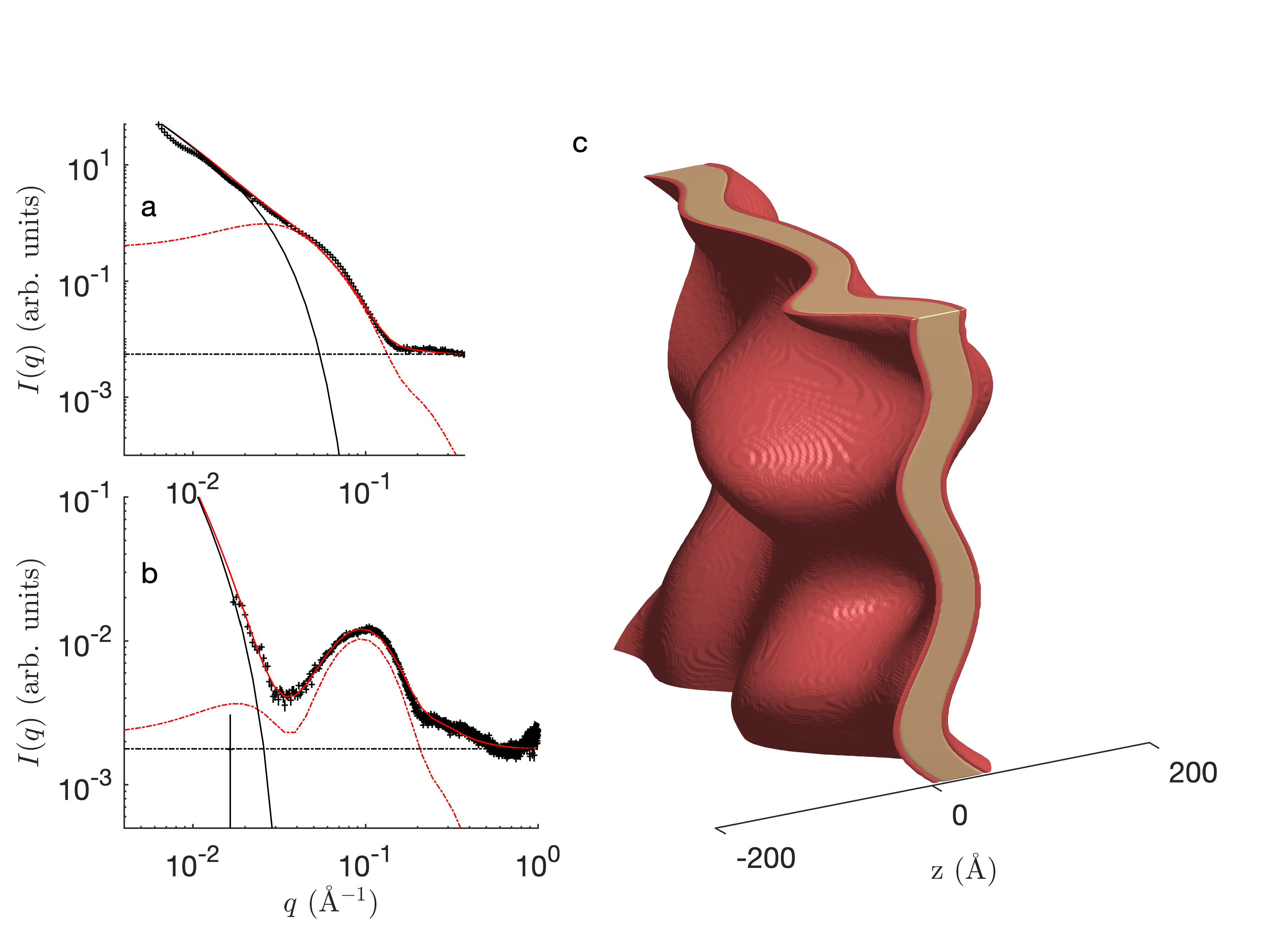}
\caption{Fitting of the SANS (a) and SAXS (b) data of the phospholipid vesicles (same as in Fig. \ref{fig:data}) with the Gaussian membrane model. The dots are the data, and the solid red line is the fitted model. The solid black line is the scattering from the average structure, and the dashed red line is the contribution from the fluctuations. The horizontal dashed black line is the background. A particular realization of the fitted model is shown in c.}
\label{fig:SASfit_GRL}
\end{figure}

Figure \ref{fig:SASfit_GRL} displays the same SANS and SAXS data as in Fig. \ref{fig:data}, fitted here with the Gaussian membrane model. The fitting parameters are (i) the clipping characteristics $Z_n$'s and $l_\alpha$ (see Fig. \ref{fig:sketch}a) and (ii) the field characteristic lengths $l_{XY}$ and $l_Z$. The four $Z$'s are symmetric with respect to the origin (the membrane is arbitrarily centred on $z=0$), and are therefore only two degrees of freedom, equivalent to specifying the thicknesses of the chain and head sections $l_C$ and $l_H$. The least-square fit yields the values $Z_2 = 22$ \AA, $Z_3 = 29$ \AA \ (with $Z_0=-Z_3$ and $Z_1=-Z_2$), $l_\alpha = 38$ \AA, $l_{XY}=80$ \AA, and $l_Z=233$ \AA. A realization of the model with those sets of parameters is displayed in Fig. \ref{fig:SASfit_GRL}c. The overall amplitude of the fluctuations (controlled by parameter $l_\alpha$) is found to be of the order of 10 nm, which is globally in line with independently-determined values for vesicles.\cite{Monzel:2016}

It is interesting to compare the values of $Z_2 = 22$ \AA \ and $Z_3-Z_2 = 7$ \AA \ to the slightly smaller values $l_C = 17$ \AA \ and $l_H = 4.5$ \AA \ inferred from the deterministic slab model (see Tab. \ref{tab:slab_fit}). To make a meaningful comparison, however, one has to note that the values of $Z$'s control the volumes of the layers and not directly their thicknesses. The volume of a layer can indeed be calculated as the integral of its one-point probability function defined in Eq. (\ref{eq:S_n}), namely
\begin{equation}
\frac{V_n}{A} = \int_{-\infty}^{+\infty} S^{(1)}_n(z) \textrm{d}z
\end{equation}
which is here expressed per unit of area of the projected membrane $A$, orthogonal to $z$. Based on Eq. (\ref{eq:Lambda1}) the relation is found to be simply $V_n/A = Z_{n}-Z_{n-1}$. In other words, increasing the amplitude of the membrane fluctuations through parameter $l_\alpha$ does not modify the membrane volume. However, the average thickness has to be calculated by dividing the volume by the actual surface area of the distorted membrane, which can be significantly larger than $A$. The average thickness is therefore evaluated as
\begin{equation}
\langle l_n \rangle = \frac{Z_n-Z_{n-1}}{a_A}
\end{equation}
where $a_A$ is the dimensionless roughness factor calculated in Sec. \ref{sec:model}. With the values of the fitted parameters, the roughness factor is $a_A \simeq 1.35$, shown as a red dot in Fig. \ref{fig:area}. With this correction, the values from the Gaussian membrane fit are $\langle l_C \rangle = 16.3$ \AA \ and $\langle l_H \rangle = 5.2$ \AA, in fair agreement with the flat slab model.

In addition to the average thickness, another important characteristic of the membrane is its compressibility, which can be quantified as the variability (in time and space) of the membrane thickness. In the case where $l_\alpha$ is smaller than $l_Z$, as is the case for the structure in Fig. \ref{fig:SASfit_GRL}, the position $z_n$ of the $n^{th}$ interface is approximately a Gaussian variable with mean value $Z_n$ and standard deviation $l_\alpha$. Moreover, the correlation between the fluctuations of the $n^{th}$ and $m^{th}$ interfaces is $g_W(r_{xy},r_z=Z_m-Z_n,\tau)$. The variance of the thickness fluctuation - for one given position in the $xy$ plane $r_{xy}=0$, and one given time $\tau = 0$ - is therefore obtained by considering the difference of two correlated Gaussian variables, namely
\begin{equation} 
\langle (z_n-z_m)^2 \rangle = (Z_n-Z_m)^2 + 2 l_\alpha^2 \left[ 1 - g_W(r_{xy}=0, r_z=Z_n-Z_m, \tau=0) \right]
\end{equation} 
where the first and second terms account for the average and fluctuating contributions. Using the expression in Eq. (\ref{eq:gW}) for the field correlation function $g_W$ then provides the following expression for the standard deviation of the thickness fluctuation relative to the mean
\begin{equation} \label{eq:compressibility}
\frac{\sigma_{z_n-z_m}}{Z_n-Z_m} \simeq \sqrt{2} \ \frac{l_\alpha}{l_Z}
\end{equation}
in the limit where $|Z_n-Z_m| \ll l_Z$. Using the values of $l_\alpha$ and $l_Z$ from the SAXS and SANS fit, one obtains the numerical value $\sigma_{z_n-z_m}/(Z_n-Z_m) \simeq 0.23$. This compares resonably with the 15 \% polydispersity of the slab model fitted in Fig. \ref{fig:data}, but the interpretation is different here. In the slab model, one assumes that all the membranes have permanently slightly different thicknesses. In the Gaussian model, all the membranes are indeed identical but they fluctuate, also in thickness. Because SAXS and SANS capture only an instantaneous snapshot of a fluctuating structure, they cannot discriminate between the two scenarios. 

The fluctuations, however, are central to the inelastic scattering underlying the neutron spin echo (NSE) data in Fig. \ref{fig:data}a. In classical approaches for membrane scattering data analysis, the elastic (SANS, SAXS) and inelastic (NSE) data are analyzed independently of one another. Elastic scattering data are typically analyzed through static models, which can be structurally sophisticated \cite{Pabst:2000,Heberle:2013,Chakraborty:2020,Chappa:2021,Lewis-Laurent:2021} and possibly account also for the diffuse scattering from membrane deformation.\cite{Hamley:2022} In classical approaches, the inelastic scattering data is analyzed independently of the SANS or SAXS, {\it e.g.} through a Zilman-Granek approach \cite{Zilman:1996,Zilman:2002} whereby the intermediate scattering function is modelled as stretched exponentials at given $q$'s, with parameters dependent on the elastic moduli and viscosities of the membrane constituents.\cite{Nagao:2023} The present Gaussian random layer model offers the possibility of building the inelastic scattering data analysis on the results of the SAXS or SANS. 

\begin{figure}
\centering
\includegraphics[width=14cm]{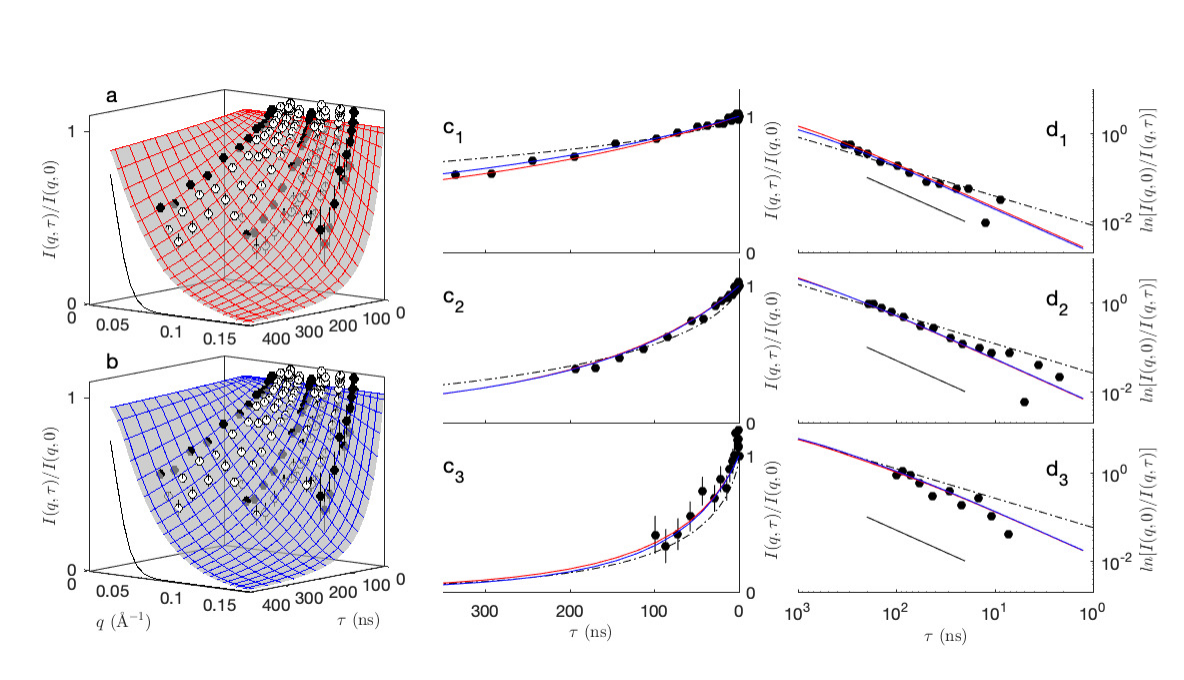}
\caption{One-parameter fits of the Neutron Spin Echo data with the Gaussian membrane model, for the exponential (a, $\tau_c=925$ ns) and diffusive (b, $D=1.65$ \AA$^{2}$/ns) time dependences. The same data are reproduced for clarity in c$_1$ ($q = 0.046 $ \AA$^{-1}$), c$_2$  ($q = 0.082 $ \AA$^{-1}$), and c$_3$  ($q = 0.122 $ \AA$^{-1}$), together with the exponential (red) or diffusive (blue) dynamic models. Panels d$_1$ to d$_3$ are the same data on logarithmic scales to highlight stretched exponentials; the solid black line is a standard exponential. The dashed black lines is the Zilman-Granek model (same fit as in Fig. \ref{fig:data}a). The solid black lines shown at $\tau\simeq 400$ ns in a and b is the asymptotic value $I(q,\tau \to \infty)/I(q,0)$.}
\label{fig:NSEfit}
\end{figure}

\begin{figure}
\centering
\includegraphics[width=16cm]{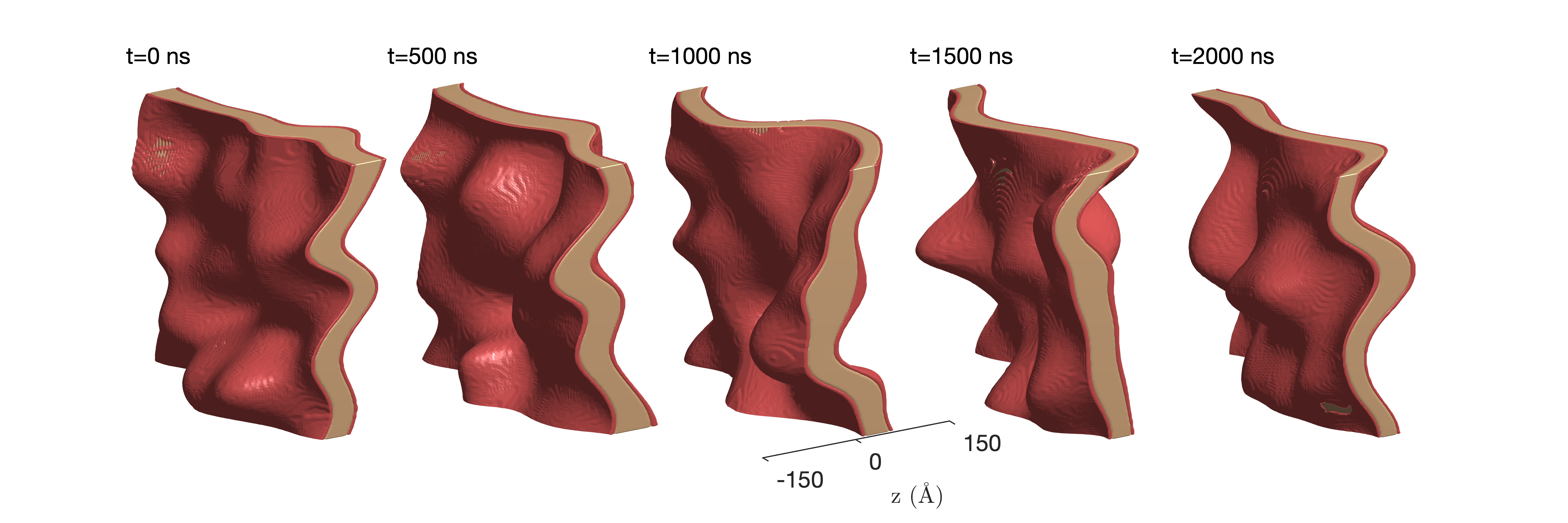}
\caption{Time-dependent realization of the Gaussian random layer model, with parameters fitted from the SAXS and SANS data (Fig. \ref{fig:SASfit_GRL}c), and exponential time-dependence of $g_W(\mathbf{r},\tau)$ with correlation time from the NSE data (Fig. \ref{fig:NSEfit}). }
\label{fig:realization_vs_time}
\end{figure}

Figure \ref{fig:NSEfit} compares the NSE data from Fig. \ref{fig:data}c (dots) with the intermediate scattering function calculated from the Gaussian membrane model. We consider only the two non-differentiable dynamics, corresponding to Figs. \ref{fig:GRF_time}a and \ref{fig:GRF_time}c. For the calculations, the values of the structural parameters inferred from the SAXS and SANS are kept constant; the only additional and fitting parameter is the one describing the time-dependence of the Gaussian field through the $\tau$ dependence of the correlation function $g_W(\mathbf{r},\tau)$. Two models are considered here, namely: the exponential correlation function (Eqs. \ref{eq:g_separable} and \ref{eq:f_exp}) parameterized through the correlation time $\tau_c$, and the diffusive model (Eq. \ref{eq:g_diffusive}) parameterized through diffusion coefficient $D$. The two models fit equally well the NSE data with either $\tau_c = 925$ ns or $D \simeq 1.65$ \AA$^2$/ns, as shown in Figs. \ref{fig:NSEfit}a and \ref{fig:NSEfit}b, respectively. To compare the two models, one has to notice that the correlation time of the diffusive Gaussian field can be estimated as the time needed for a two-dimensional random walker to travel a distance comparable with the correlation length $l_{XY}$, {\it i.e.} as $l_{XY}^2/(4D) \simeq 970$ ns. Expectedly, this value is comparable with $\tau_c$ obtained from the fit of the exponential correlation. A time-dependent realization of the exponential model is shown in Fig. \ref{fig:realization_vs_time}, over a duration of approximately $2 \tau_c$. Note how both the bending and thickness fluctuations are captured in the latter realizations.

Figures \ref{fig:NSEfit}c$_1$ to \ref{fig:NSEfit}c$_3$ compare the fits of the two Gaussian membrane models with that of the Zilman-Granek model in the form of Eq. (\ref{eq:ZG}), {\it i.e.} with a single fitting parameter $B$. To understand the qualitative differences between the models, the same data are plotted again in Figs. \ref{fig:NSEfit}d$_1$ to \ref{fig:NSEfit}d$_3$ in the form of $\log[ - \log[I(q,\tau)/I(q,0)]]$ {\it versus} $\log(\tau)$ in order for stretched exponentials to appear as straight lines. On such plots, it appears that the intermediate scattering function of the Gaussian membrane follows the same stretched-exponential of $\tau$ as the Zilman-Granek model (with exponent $2/3$) only for large $\tau$ (see {\it e.g.} Fig. \ref{fig:NSEfit}d$_3$). For short times $\tau$ the dependence is closer to a classical exponential (with exponent 1), which incidentally seems to describe the NSE data better. The two smooth models considered in Sec. \ref{sec:time} - the hyperbolic secant correlation (Fig. \ref{fig:GRF_time}b) and the ballistic model (Fig. \ref{fig:GRF_time}d) - lead to quadratic scaling $I(q,\tau) \simeq \exp(-\tau^2)$ for short $\tau$ (not shown), and they are therefore not suitable for the NSE data analysis. This is in line with earlier observations with emulsion scattering.\cite{Gommes:2021}

In the context of NSE studies, the question of membrane compressibility is often addressed through head-contrast experiments, which exhibit enhanced dynamics at intermediate values of $q$.\cite{Nagao:2009,Woodka:2012} Such data are usually analyzed by fitting the intermediate scattering function with a stretched exponential $I(q,\tau)/I(q,0) = \exp[-(\Gamma \tau)^{2/3}]$ to extract a $q$-dependent relaxation rate $\Gamma(q)$. When plotting the results as $\Gamma(q)/q^3$ -  to visualize deviations from the $q^3$ scaling of the Zilman-Granek theory - the data often exhibit a peak, which is considered as a signature of thickness fluctuations. \cite{Chakraborty:2020,Nagao:2023,Luo:2023} The present Gaussian model offers the possibility of exploring this data analysis procedure, based on an a mathematically exact model.

\begin{figure}
\centering
\includegraphics[width=8cm]{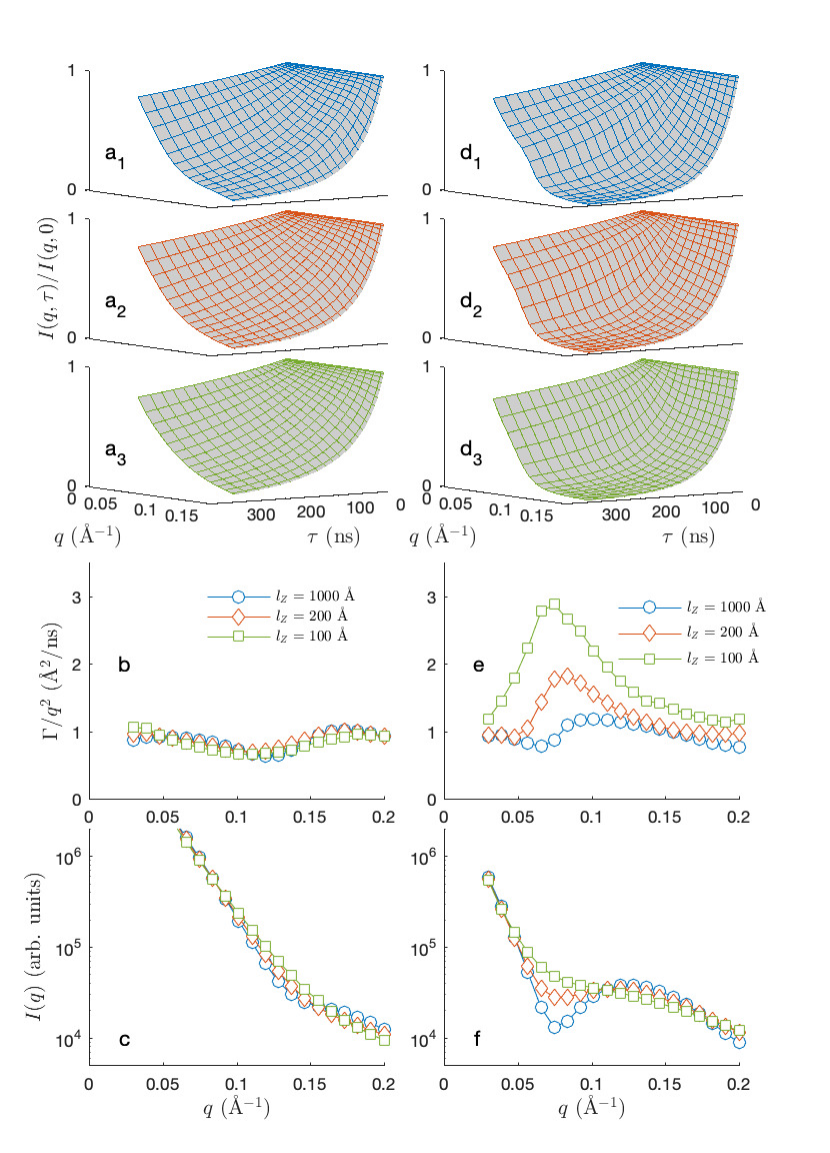}
\caption{Intermediate scattering functions (a and d) and SANS intensities (c and f) calculated from the Gaussian model with the same parameter as inferred from the SANS and NSE fits, only with different values of the correlation length $l_Z$, equal to 1000 \AA \ (a$_1$ and d$_1$), 200 \AA \ (a$_2$ and d$_2$) and 100 \AA \ (a$_3$ and d$_3$). The left column is for chain contrast (a to c) and the right column is for head contrast (d to f). The corresponding relaxation rates are plotted in b and e.}
\label{fig:NSE_compressibility}
\end{figure}

Figure \ref{fig:NSE_compressibility} plots the intermediate scattering function of the Gaussian membrane model for both chain and head contrasts, using the same parameters as inferred from the SANS and NSE fits in Figs. \ref{fig:SASfit_GRL} and \ref{fig:NSEfit}, only with different values of the field correlation length $l_Z$ perpendicular to the membrane. The latter parameter controls the thickness fluctuations of the membrane through Eq. (\ref{eq:compressibility}): the values used in the figure are $l_Z=1000$ \AA, 200 \AA \ and 100 \AA, corresponding to 5 \%, 25 \% and 50 \% thickness fluctuations, respectively. Because the intermediate scattering function is well described by a simple exponential for small $\tau$ (see Fig. \ref{fig:NSEfit}d$_1$ to \ref{fig:NSEfit}d$_3$), the relaxation rate $\Gamma(q)$ was evaluated by fitting the calculated values as $I(q,\tau)/I(q,0) \simeq \exp[-\Gamma \tau]$ and the values are reported in Fig. \ref{fig:NSE_compressibility}b and \ref{fig:NSE_compressibility}e. In the case of chain contrast the relaxation rate follows an overall scaling close to $q^{2}$, with no visible impact of the thickness fluctuation. In that respect, the Gaussian membrane model exhibits a typical diffusion phenomenology with relaxation rate scaling linearly with $\tau$ and quadratically with $q$. The case of head contrast is distinctly different; the relaxation rate is found to exhibit a peaked maximum, which is clearly correlated with thickness fluctuations. The $q$ position of the maximum approximately corresponds to the minimum of the SANS intensity (see Fig. \ref{fig:NSE_compressibility}f).

\section{Conclusions}

We have developed a mathematical model of a fluctuating membrane, for the purpose of jointly analyzing small-angle scattering of x-rays and/or neutrons, as well as neutron spin-echo data, with arbitrary contrasts, within a single theoretical framework. The model builds on a non-homogeneously clipped time-dependent Gaussian random field, which modelling procedure provides one with general analytical expressions for the intermediate scattering function. The availability analytical results significantly reduces the computational burden of data analysis; all analyses and simulations presented in this paper were conducted on a laptop computer.

Our Gaussian membrane model realistically captures the structure and dynamics of fluctuating membranes with a handful of meaningful parameters. Although the model is descriptive, exploring the influence of its parameters provides direct insight into the physics of elastic and inelastic scattering by membranes. In particular, our analysis enabled us to understand how membrane compressibility contributes to forward elastic scattering. It also confirmed through exact mathematical calculations how compressibility significantly enhances the relaxation rates in neutron spin echo data, in the case of head-contrast only. 

Moreover the synthetic description of membrane structure and dynamics, enables one to comprehensively analyze large datasets comprising both elastic and inelastic scattering patterns measured for arbitrary contrasts with one and a single model. All the structural and dynamic information is then condensed in a coherent and robust way into a few meaningful parameters. This contrasts with classical approaches in this context, where elastic and inelastic scattering patterns are traditionally analyzed independently of one another with data-specific models. Analyzing jointly different datasets, also offers the prospect of extracting meaningful information that is not present in the individual datasets when they are considered separately.

We illustrated our approach with the analysis of SAXS, SANS and NSE data measured at 37 $^{\circ}$C on vesicles prepared from porcine brain phospholipids, close to physiological conditions. The elastic scattering data were analyzed first, to identify the structural parameters. The latter are equivalent to (i) the thicknesses of the hydrophobic chains and hydrophilic heads of the membrane molecules, (ii) the amplitude of the membrane deformations as well as (iii) their lateral sizes, and (iv) the thickness fluctuation. The latter three parameters are absent from classical membrane models, but they can be inferred from the joint SAXS and SANS analysis. The membranes investigated in this work are found to undergo bending fluctuations with amplitude around 10 nm and thickness fluctuations of the order of 8 \AA.  

Building on these values of the structural parameters, the complete NSE dataset was fitted with a single additional parameter characterizing the dynamics of the fluctuations. The two specific dynamic models we considered fit equally well the data. Interestingly, they both predict a $\tau^{2/3}$ scaling of the relaxation rate (similar to the Zilman-Granek model) for asymptotically large times. At finite times, however, the calculated relaxation rate is diffusion-like and it scales linearly with $\tau$ and quadraticaly with $q$. Our data incidentally exhibit this specific type of scaling. 

\begin{acknowledgments}
CJG is grateful to the Funds for Scientific Research (F.R.S.-FNRS, Belgium) for a Research Associate position, and for supporting part of this work through grant PDR T.0100.22.
\end{acknowledgments}


%

\end{document}


\title[]{Supporting Information to \\ \vspace{1cm} A Gaussian model of fluctuating membrane and its scattering properties}

\author{Cedric J. Gommes}  \email{cedric.gommes@uliege.be}
\affiliation{Department of Chemical Engineering, University of Li\`ege B6A, All\'ee du Six Ao\^ut 3, 4000 Li\`ege, Belgium}
\author{Purushottam S. Dubey} 
\affiliation{Forschungszentrum J\"ulich GmbH, J\"ulich Center for Neutron Science at the Heinz Maier Leibnitz Zentrum, Lichtenbergstrasse 1, 85747 Garching, Germany}
\author{Andreas M. Stadler}
\affiliation{Forschungszentrum J\"ulich GmbH, J\"ulich Center for Neutron Science, 52425 J\"ulich, Germany}
\affiliation{Institute of Physical Chemistry, RWTH Aachen University, Landoltweg 2, 52056 Aachen, Germany}
\author{Baohu Wu}
\affiliation{Forschungszentrum J\"ulich GmbH, J\"ulich Center for Neutron Science at the Heinz Maier Leibnitz Zentrum, Lichtenbergstrasse 1, 85747 Garching, Germany}
\author{Orsolya Czakkel}
\affiliation{Institut Laue-Langevin, 71 avenue des Martyrs, 38042 Grenoble Cedex 9, France}
\author{Lionel Porcar}
\affiliation{Institut Laue-Langevin, 71 avenue des Martyrs, 38042 Grenoble Cedex 9, France}
\author{Sebastian Jaksch}
\affiliation{Forschungszentrum J\"ulich GmbH, J\"ulich Center for Neutron Science at the Heinz Maier Leibnitz Zentrum, Lichtenbergstrasse 1, 85747 Garching, Germany}
\affiliation{European Spallation Source (ESS) ERIC, Partikelgatan 2, 224 84 Lund, Sweden}
\author{Henrich Frielinghaus}
\affiliation{Forschungszentrum J\"ulich GmbH, J\"ulich Center for Neutron Science at the Heinz Maier Leibnitz Zentrum, Lichtenbergstrasse 1, 85747 Garching, Germany}
\author{Olaf Holderer} \email{o.holderer@fz-juelich.de}
\affiliation{Forschungszentrum J\"ulich GmbH, J\"ulich Center for Neutron Science at the Heinz Maier Leibnitz Zentrum, Lichtenbergstrasse 1, 85747 Garching, Germany}

\date{\today}

\maketitle

\renewcommand{\theequation}{SI-\arabic{equation}} 
\setcounter{equation}{0}  

\renewcommand{\thefigure}{SI-\arabic{figure}} 
\setcounter{figure}{0}  

\section{Specific surface area of the interfaces}

All interfaces in the stacked layer model of the main text (Fig. 2) are defined in the exact same mathematical way, only with a shifting along direction $z$. As a consequence, all interfaces have the same (average) surface area. We here derive a general expression for the specific surface area of the interface, as a function of model parameters $l_{xy}$ and $l_z$ (for the Gaussian field) as well as of the clipping slope $l_\alpha$. The area is expressed as the geometrical area of the distorted interface, per unit area of the projected plane orthogonal to $z$. This is mathematically equivalent to a dimensionless roughness factor, larger than one.

To calculate the area of the interfaces in the model of main text it is sufficient to consider a two-phase structure, defined as
\begin{equation}
\mathcal{I}_0(\mathbf{x}) = H\left( W(\mathbf{x} - \alpha(\mathbf{x})\right) \quad \textrm{and} \quad \mathcal{I}_1(\mathbf{x}) = 1 - \mathcal{I}_0(\mathbf{x})
\end{equation}
The two-point probability function is 
\begin{equation}
S^{(2)}_{01}(\mathbf{x_1},\mathbf{x}_2) = \Lambda_1[\alpha(\mathbf{x}_1)] - \Lambda_2[\alpha(\mathbf{x}_1),\alpha(\mathbf{x}_2),g_W(\mathbf{x}_1-\mathbf{x}_2)]
\end{equation}
where the time-dependence of $g_W$ is irrelevant here, as all calculations here are meant for $\tau=0$.

General expressions for the specific surface area of non-stationary anisotropic structures are derived in the Supporting Information of Ref. \onlinecite{Gommes:2009}. In particular, the specific area of the interface between, say, phases 0 and 1 at any point $\mathbf{x}$ is related to the two-point probability function via
\begin{equation} \label{eq:a}
\frac{a_{01}(\mathbf{x})}{4} = \frac{1}{4 \pi} \int \textrm{d} \hat \omega \lim_{r \to 0} \frac{\partial}{\partial r} S_{01}^{(2)}(\mathbf{x},\mathbf{x} + r \hat \omega)
\end{equation}
where $\hat \omega$ is a unit vector, and the integral is over the unit sphere. This is Eq. B1 of Ref. \onlinecite{Gommes:2009}, which we have rewritten here in terms of a cross-correlation function.

Based on Eq. B17 of Ref. \onlinecite{Gommes:2009}, the derivative of the two-point function can be written as
\begin{align} \label{eq:dS2}
\lim_{r \to 0} \frac{\partial}{\partial r} & S^{(2)}_{01}(\mathbf{x},\mathbf{x}+ r \hat \omega) = \frac{\hat \omega \cdot \nabla \alpha}{2 \sqrt{2 \pi}} e^{-\frac{\alpha^2}{2}} \cr
& + \frac{1}{l_{\hat \omega}} \frac{1}{\pi \sqrt{2}} e^{-\alpha^2/2}  \left\{ \exp\left[ -\frac{[l_{\hat \omega} \hat \omega \cdot \nabla \alpha]^2}{4}\right]
+ \sqrt{\pi} \frac{|l_{\hat \omega} \hat \omega \cdot \nabla \alpha|}{2} \textrm{erf} \left[ \frac{| l_{\hat \omega} \hat \omega \cdot \nabla \alpha|}{2}\right]
 \right\}
\end{align}
where $l_{\hat \omega}$ is the characteristic length of the Gaussian field in direction $\hat \omega$, defined so that
\begin{equation}
g_W(r \hat \omega) = 1 - \left( \frac{r}{l_{\hat \omega}} \right)^2+ \ldots
\end{equation}
for asymptotically small values of $r$. 

In the case of the main text, with anisotropic field with in-plane and out-of-plane correlation lengths $l_{XY}$ and $l_Z$, this is written as
\begin{equation}
\frac{1}{l_{\hat \omega}} = \sqrt{\frac{\sin^2(\theta)}{l_{XY}^2} + \frac{\cos^2(\theta)}{l_{Z}^2}} = \frac{1}{l_{XY}} \sqrt{1 +\left[ \left(\frac{l_{XY}}{l_Z}\right)^2 -1\right] \cos^2(\theta)}
\end{equation}
where $\theta$ is the azimuthal angle with direction $Z$. Moreover, because the clipping constant only depends on the $Z$ coordinate, one has $\nabla \alpha = \mathbf{e}_z/l_\alpha$. In that case the orientation-dependence of $l_{\hat \omega} \hat \omega \cdot \nabla \alpha$ is the following
\begin{equation}
l_{\hat \omega} | \hat \omega \cdot \nabla \alpha  | = \frac{(l_{XY}/l_\alpha) \cos(\theta)}{ \sqrt{1 +\left[ (l_{XY}/l_Z)^2 -1\right] \cos^2(\theta)}}
\end{equation}
To calculate the specific surface area of the interface, one has to (i) calculate the rotational average of the right-hand-side of Eq. (\ref{eq:dS2}) and (ii) insert the result into Eq. (\ref{eq:a}). At this stage, one can already check that in the case where the clipping threshold is constant, the right-hand side of Eq. (\ref{eq:dS2}) reduces to $\exp(-\alpha^2/2)/(\pi \sqrt{2} l_\omega)$. If, additionally, the Gaussian field is isotropic, Eq. (\ref{eq:a}) provides the classical result
\begin{equation} \label{eq:a_classical}
a_{01} = \frac{2^{3/2}}{\pi l_W} \exp(-\alpha^2/2)
\end{equation}
as it should, for the specific surface area of the homogeneously-clipped isotropic Gaussian random field \cite{Teubner:1991}. 

To keep the same formal equation as in Eq. (\ref{eq:a_classical}) we write the general result as
\begin{equation} \label{eq:a_classical}
a_{01} = \frac{2^{3/2}}{\pi l_{XY}} \exp(-\alpha^2/2) \times F
\end{equation}
where $F$ is a function of $l_Z/l_{XY}$, and $l_\alpha/l_{XY}$ calculated as 
\begin{align}
F = \int_0^1 \textrm{d}\mu & \left\{1+\left[ \left(\frac{l_{XY}}{l_Z}\right)^2-1 \right]\mu^2 \right\}^{1/2} \cr
&\times \left\{ \exp\left[ -\frac{[l_{\hat \omega} \hat \omega \cdot \nabla \alpha]^2}{4}\right] \right.
+ \left. \sqrt{\pi} \frac{|l_{\hat \omega} \hat \omega \cdot \nabla \alpha|}{2} \textrm{erf} \left[ \frac{| l_{\hat \omega} \hat \omega \cdot \nabla \alpha|}{2}\right]
\right\}
\end{align}

\newpage 

\section{Derivation of Eq. (26) of the main text}

The fluctuation contribution to the scattering cross section is calculated as
\begin{equation} \label{eq:I}
\tilde I(\mathbf{q},\tau) = \sum_{n=0}^{N+1} \sum_{m=0}^{N+1} b_n b_m \tilde P_{m,n}(\mathbf{q},\tau)
\end{equation}
where $\tilde P_{m,n}(\mathbf{q})$ is the following Fourier transform
\begin{equation} \label{eq:Pmn_3D}
\tilde P_{m,n}(\mathbf{q},\tau) = \int \textrm{d}V_1 \int \textrm{d}V_2 \quad e^{-i \mathbf{q} \cdot \left(\mathbf{x}_1 -  \mathbf{x}_2 \right) } 
\tilde S^{(2)}_{m,n}(\mathbf{x_1},\mathbf{x_2},\tau)
\end{equation}
and
\begin{equation}
\tilde S^{(2)}_{m,n}(\mathbf{x_1},\mathbf{x_2},\tau) = S^{(2)}_{m,n}(\mathbf{x_1},\mathbf{x_2},\tau) - S^{(1)}_m(\mathbf{x}_1) S^{(1)}_m(\mathbf{x}_1)
\end{equation}
is the two-point probability function, from which the asymptotic value (for large $\tau$) is subtracted.

To take advantage of the symmetry of the layer model, it is convenient to make a change of variable $\mathbf{r} = \mathbf{x}_2-\mathbf{x}_1$ in Eq. (\ref{eq:Pmn_3D}), and to integrate on $\mathbf{x}_1$. This boils down to introducing the correlation function
\begin{equation} \label{eq:C}
\tilde C_{m,n}(\mathbf{r},\tau) = \int_{-\infty}^{+\infty} \tilde S^{(2)}_{m,n}(z \mathbf{e}_z, z \mathbf{e}_z + \mathbf{r},\tau) \textrm{d}z
\end{equation}
where $\mathbf{e}_z$ is a unit vector orthogonal to the layer. With this notation, the scattering cross section {\it per unit area of the membrane} is written as the 3D Fourier transform of 
\begin{equation} \label{eq:Cb}
\tilde C_b(\mathbf{r},\tau) = \sum_{n=0}^{N+1} \sum_{m=0}^{N+1} b_n b_m \tilde C_{m,n}(\mathbf{r},\tau)
\end{equation}
where the linear combination is identical to Eq. \ref{eq:I}.

Quite generally, the two-point probability function of clipped Gaussian-filed models are conveniently expressed in terms of bivariate error functions $\Lambda_2[\alpha_1,\alpha_2,g]$, defined as the probability for two correlated Gaussian variables (with correlation coefficient $g$) to be larger than $\alpha_1$ and $\alpha_2$, respectively. In principle, the error function $\Lambda_2$ can be evaluated directly as the following two-dimensional integral
\begin{equation}
\Lambda_{2}[\alpha_1,\alpha_2,g] = \frac{1}{2 \pi \sqrt{1-g^2}} \int_{\alpha_1}^\infty \textrm{d}x \int_{\alpha_2}^\infty \textrm{d}y \quad 
e^{-(x^2+y^2 -2 g xy)/2}
\end{equation}
where the integrand is the bivariate Gaussian distribution. In practice, a mathematically equivalent yet faster alternative consists in calculating \cite{Berk:1991,Teubner:1991}
\begin{equation} \label{eq:Lambda2_exact}
\Lambda_{2}[\alpha_1,\alpha_2,g] = \Lambda_{1}[\alpha_1] \Lambda_{1}[\alpha_2]
+ \frac{1}{2 \pi} \int_0^{\textrm{asin}(g)} \textrm{d}\theta \ e^{-\frac{\alpha_1^2 + \alpha_2^2 - 2 \alpha_1 \alpha_2 \sin(\theta) }{2 \cos^2(\theta)}}
\end{equation}
which only requires a one-dimensional integral, and is numerically easier to handle. In Eq. (\ref{eq:Lambda2_exact}), the function $\Lambda_1[]$ is the univariate error function defined in Eq. (24) of the main text.

\subsection{Two-point probability functions $S_{m,n}$}
\label{sec:Smn}

\subsubsection{Self-correlations, $S_{n,n}$} 

The self-correlation of the $n^{th}$ lamella $S_{n,n}(\mathbf{x}_1,\mathbf{x}_2,\tau)$ is 
\begin{equation}
S_{n,n}(\mathbf{x}_1,\mathbf{x}_2,\tau) = \langle \mathcal{I}_n(\mathbf{x}_1,t)  \mathcal{I}_n(\mathbf{x}_2,t+\tau) \rangle
\end{equation}
In the case of the indicator functions defined in Eq. (10)  of the main text, this is calculated as
\begin{align}
S_{n,n}(\mathbf{x}_1,\mathbf{x}_2,\tau) = \Big< \left( H\left[ W(\mathbf{x}_1,t_1) - \alpha_n(\mathbf{x}_1) \right] - H\left[ W(\mathbf{x}_1,t_1) - \alpha_{n-1}(\mathbf{x}_1) \right] \right) \cr
 \times \left( H\left[ W(\mathbf{x}_2,t_2) - \alpha_n(\mathbf{x}_2) \right] - H\left[ W(\mathbf{x}_2,t_2) - \alpha_{n-1}(\mathbf{x}_2) \right] \right) \Big>
\end{align}
In terms of the bivariate error function $\Lambda_2$, this can which leads to
\begin{align}
S_{n,n}(\mathbf{x}_1,\mathbf{x}_2,\tau) &= \Lambda_2 \left[ \alpha_n(\mathbf{x}_1), \alpha_n(\mathbf{x}_2), g_{12} \right]  
- \Lambda_2 \left[ \alpha_n(\mathbf{x}_1), \alpha_{n-1}(\mathbf{x}_2), g_{12} \right] \cr
&- \Lambda_2 \left[ \alpha_{n-1}(\mathbf{x}_1), \alpha_{n}(\mathbf{x}_2), g_{12} \right] 
+ \Lambda_2 \left[ \alpha_{n-1}(\mathbf{x}_1), \alpha_{n-1}(\mathbf{x}_2), g_{12} \right] 
\end{align}
with
\begin{equation}
g_{12} = g_W\left( \mathbf{x}_1-\mathbf{x}_2, \tau \right)
\end{equation}
where $g_W$ is the field correlation function, defined in Eq. (4) of the main text.

In the case of the outer regions of the layer stack, the indicator functions $\mathcal{I}_0$ and $\mathcal{I}_{N+1}$ are defined differently than the inner layers , which leads to
\begin{equation} \label{eq:S_0_0}
S_{0,0} (\mathbf{x}_1,\mathbf{x}_2,\tau) = \Lambda_2\left[ \alpha_0(\mathbf{x}_1), \alpha_0(\mathbf{x}_2), g_{12} \right]
\end{equation}
and 
\begin{align} \label{eq:S_N+1_N+1}
S_{N+1,N+1} (\mathbf{x}_1,\mathbf{x}_2,\tau) &= 1 - \Lambda_1\left[ \alpha_N(\mathbf{x}_1)\right]  - \Lambda_1\left[ \alpha_N(\mathbf{x}_2)\right]  \cr
&+ \Lambda_2\left[ \alpha_N(\mathbf{x}_1), \alpha_N(\mathbf{x}_2), g_{12} \right]
\end{align}

\subsubsection{Cross-correlations, $S_{m,n}$ with $m \neq n$} 

In the case of $S_{n,m}$ with $1 \le n \le N$ and $1 \le m \le N$, the same procedure as earlier leads to
\begin{align}
S_{n,m}(\mathbf{x}_1,\mathbf{x}_2,\tau) & = \Lambda_2 \left[ \alpha_n(\mathbf{x}_1), \alpha_m(\mathbf{x}_2), g_{12}) \right] 
 - \Lambda_2 \left[ \alpha_n(\mathbf{x}_1), \alpha_{m-1}(\mathbf{x}_2), g_{12} \right] \cr
& - \Lambda_2 \left[ \alpha_{n-1}(\mathbf{x}_1), \alpha_{m}(\mathbf{x}_2), g_{12} \right] 
+ \Lambda_2 \left[ \alpha_{n-1}(\mathbf{x}_1), \alpha_{m-1}(\mathbf{x}_2), g_{12} \right]
\end{align}
which incidentally applies also if $m=n$.

The case where either $n$ or $m$ is equal to 0 or $N+1$, requires a specific expression, namely
\begin{equation}
S_{0,n}(\mathbf{x}_1,\mathbf{x}_2,\tau) = 
\Lambda_2 \left[ \alpha_0(\mathbf{x}_1), \alpha_{n}(\mathbf{x}_2), g_{12} \right] - \Lambda_2 \left[ \alpha_{0}(\mathbf{x}_1), \alpha_{n-1}(\mathbf{x}_2), g_{12} \right] 
\end{equation}
\begin{align}
S_{n,N+1}(\mathbf{x}_1,\mathbf{x}_2,\tau) & = 
\Lambda_1\left[ \alpha_n(\mathbf{x}_1)\right]  - \Lambda_1\left[ \alpha_{n-1}(\mathbf{x}_2)\right] -\Lambda_2 \left[ \alpha_n(\mathbf{x}_1), \alpha_{N}(\mathbf{x}_2), g_{12} \right] \cr
&+ \Lambda_2 \left[ \alpha_{n-1}(\mathbf{x}_1), \alpha_{N}(\mathbf{x}_2), g_{12} \right] 
\end{align}
for $1 \leq n \leq N$, and 
\begin{align} \label{eq:S_0_N+1}
S_{0,N+1}(\mathbf{x}_1,\mathbf{x}_2,\tau) & = 
\Lambda_1\left[ \alpha_0(\mathbf{x}_1)\right]   -\Lambda_2 \left[ \alpha_0(\mathbf{x}_1), \alpha_{N}(\mathbf{x}_2), g_{12} \right] \cr
\end{align}

\subsubsection{Correlation function $\tilde C_{m,n}(\mathbf{r},\tau)$} 

In Sec. \ref{sec:Smn}, the two point functions $S_{m,n}$ are obtained as linear combinations of $\Lambda_2$ functions, with additive contributions that account for the average  values. The fluctuation contribution to the two-point functions $\tilde S_{m,n}$ is obtained by removing all additive contributions (independent of $g_{12}$) and replacing $\Lambda_2$ by $\tilde \Lambda_2$ defined as
\begin{equation} \label{eq:tilde_Lambda2}
\tilde \Lambda_{2}[\alpha_1,\alpha_2,g] = 
\frac{1}{2 \pi} \int_0^{\textrm{asin}(g)} \textrm{d}\theta \ e^{-\frac{\alpha_1^2 + \alpha_2^2 - 2 \alpha_1 \alpha_2 \sin(\theta) }{2 \cos^2(\theta)}}
\end{equation}
Contrary to $\Lambda_2$, the function $\tilde \Lambda_2$ vanishes for $g_{12}=0$ independently of $\alpha_1$ and $\alpha_2$. 

From Eq. \ref{eq:C}, the values of $\tilde C_{m,n}(\mathbf{r})$ are then obtained as linear combinations of 
\begin{equation} \label{eq:Gamma_mn}
\Gamma_{m,n} (\mathbf{r},\tau) = \int_{-\infty}^{+\infty} \tilde \Lambda_2 \left[ \frac{z-Z_m}{l_\alpha},  \frac{z-Z_n - r_z}{l_\alpha}, g_W(\mathbf{r},\tau) \right] \ \textrm{d}z
\end{equation}
where we have have written explicitly the $z$ dependence of the clipping function $\alpha(\mathbf{x})$. Using the expression in \ref{eq:tilde_Lambda2}, the function $\Gamma_{m,n}$ can eventually be written as
\begin{equation} \label{eq:Gamma}
\Gamma_{m,n} (\mathbf{r},\tau)  = l_\alpha G \left[ \frac{|Z_m - Z_n -r_z |}{l_\alpha} , g_W(\mathbf{r},\tau) \right]
\end{equation}
with 
\begin{eqnarray}
G(\eta,g) &=& \frac{\eta}{2} \left[ \textrm{erf}\left( \frac{\eta}{2} \right) - \textrm{erf}\left( \frac{\eta}{2\sqrt{1-g}} \right)  \right] \cr
&+& \frac{1}{\sqrt{\pi}} \left[ \exp\left(-\frac{\eta^2}{4}\right) - \sqrt{1-g} \exp\left(-\frac{\eta^2}{4(1-g)} \right) \right]
\end{eqnarray}
The derivation of Eq. \ref{eq:Gamma} proceeds by (i) expressing explicitly $\tilde \Lambda_2$ in Eq. \ref{eq:Gamma_mn} through  Eq. \ref{eq:tilde_Lambda2}, (ii) inverting the order of integration of $z$ and $\theta$, and (iii) making a change of variable from $\theta$ to $t^2 = 1 + \sin(\theta)$ in the resulting integral. 

\subsection{Scattering cross section $\tilde I(\mathbf{q},\tau)$}

We derive now Eq. (26) of the main text. That expression is general, and it applies to any number of layers. We proceed here one step at a time, and consider successively an increasing number of interface in the stack of layers.

\subsubsection{One interface, $N=1$}

In the case of a single interface, $N=1$, the sum in Eq. \ref{eq:Cb} is written explicitly as
\begin{align}
\tilde C_b = b_0^2 \tilde C_{0,0} + b_1^2 \tilde C_{1,1} + b_0 b_1 \tilde C_{0,1} + b_0 b_1 \tilde C_{1,0}
\end{align}
As mentioned earlier, the expressions for $\tilde C_{m,n}$ are obtained from those of $S_{m,n}$ by discarding all constant terms and replacing $\Lambda_2[\alpha_m,\alpha_n,g]$ by $\Gamma_{m,n}$. From Eqs. \ref{eq:S_0_0}, \ref{eq:S_N+1_N+1} and \ref{eq:S_0_N+1}, one obtains
\begin{eqnarray}
\tilde C_{0,0} &=& \Gamma_{0,0} \cr
\tilde C_{1,1} &=& \Gamma_{0,0} \cr
\tilde C_{0,1} &=& - \Gamma_{0,0} \cr
\tilde C_{1,0} &=& - \Gamma_{0,0}
\end{eqnarray}
This yields
\begin{align}
\tilde C_b(\mathbf{r},\tau) &= (b_0 - b_1)^2 \Gamma_{0,0}(\mathbf{r},\tau)
\end{align}

\subsubsection{Two interfaces, $N=2$}

In the case of two interfaces ($N=2$), corresponding  to one layer.
\begin{align}
\tilde C_b &= b_0^2 \tilde C_{0,0} + b_1^2 \tilde C_{1,1} + b_2^2 \tilde C_{2,2}  + b_0 b_1 \tilde C_{0,1}
+ b_0 b_1 \tilde C_{1,0} \cr
&+ b_0 b_2 \tilde C_{0,2} + b_0 b_2 \tilde C_{2,0} + b_1 b_2 \tilde C_{1,2} 
+ b_1 b_2 \tilde C_{2,1}
\end{align}
with
\begin{eqnarray}
\tilde C_{0,0} &=& \Gamma_{0,0} \cr
\tilde C_{1,1} &=& \Gamma_{1,1} + \Gamma_{0,0} - \Gamma_{0,1} - \Gamma_{1,0}\cr
\tilde C_{2,2} &=& \Gamma_{1,1} \cr
\tilde C_{0,1} &=& \Gamma_{0,1}- \Gamma_{0,0} \quad \textrm{and} \quad  \tilde C_{1,0} = \Gamma_{1,0}- \Gamma_{0,0} \cr
\tilde C_{0,2} &=& - \Gamma_{0,1} \cr
\tilde C_{1,2} &=& \Gamma_{0,1}- \Gamma_{0,0} \quad \textrm{and} \quad \tilde C_{2,1} = \Gamma_{1,0}- \Gamma_{0,0}
\end{eqnarray}

This can be written as
\begin{align}
\tilde C_b(\mathbf{r},\tau) &=  (b_0 - b_1)^2  \Gamma_{0,0}(\mathbf{r},\tau) + (b_1 - b_2)^2  \Gamma_{1,1}(\mathbf{r},\tau) \cr
& + (b_0 - b_1) (b_1-b_2) \left(  \Gamma_{0,1}(\mathbf{r},\tau) +  \Gamma_{1,0}(\mathbf{r},\tau) \right)
\end{align}

\subsubsection{Three interfaces, $N=3$}

In the case of two layers, $N=3$.
\begin{align} \label{eq:I_layer}
\tilde C_b(\mathbf{r},\tau) &= (b_0 - b_1)^2  \Gamma_{0,0}(\mathbf{r},\tau) + (b_1 - b_2)^2  \Gamma_{1,1}(\mathbf{r},\tau) 
+ (b_2 - b_3)^2 \tilde \Gamma_{2,2}(\mathbf{q})  \cr
& + (b_0 - b_1) (b_1-b_2) \left( \Gamma_{0,1}(\mathbf{r},\tau) +  \Gamma_{1,0}(\mathbf{r},\tau)  \right) \cr
&+ (b_1 - b_2) (b_2-b_3) \left(  \Gamma_{1,2}(\mathbf{r},\tau) +  \Gamma_{2,1}(\mathbf{r},\tau)  \right) \cr
&+ (b_0 - b_1) (b_2-b_3) \left(  \Gamma_{0,2}(\mathbf{r},\tau) +  \Gamma_{2,0}(\mathbf{r},\tau) \right)
\end{align}

\subsubsection{$N$ interfaces}

The generalisation of the cases $N=1, 2, 3$ to any $N$ is straightforward. The result is
\begin{align} \label{eq:I_layer}
\tilde C_b(\mathbf{r},\tau) &= \sum_{n=0}^{N-1} (b_n - b_{n+1})^2 \Gamma_{n,n}(\mathbf{r},\tau)  \cr
& +  \sum_{m>n} (b_n - b_{n+1}) (b_m-b_{m+1}) \left( \Gamma_{n,m}(\mathbf{r},\tau) +  \Gamma_{m,n}(\mathbf{r},\tau)  \right) 
\end{align}
This is Eq. (26) of the main text.